\documentclass[aps,pre, reprint]{revtex4-1}
\usepackage{times}		
\usepackage{booktabs}
\usepackage{placeins}

\usepackage{amssymb,amstext,amsmath, epsf}
\usepackage{graphicx,graphics,epsfig,wrapfig}
\usepackage{listings}
\usepackage{float}
\usepackage{esint}

\usepackage{color}
\usepackage[english]{babel}
\usepackage[T1]{fontenc}
\usepackage[utf8]{inputenc}

\def\apj{{ApJ}}
\def\apjl{{ApJL}}

\def\mnras{{MNRAS}}

\def\prl{{PRL}}

\def\pre{{PRE}}

\def\jgr{{Journal of Geophysical Research}}
\def\apss{{Astrophysics and Space Science}}

\def\sovast{Soviet Astronomy}
\def\ssr{Space Science Reviews}
\def\solphys{Solar Physics}

\begin{document}

\title{Diffusion and Radiation in Magnetized Collisionless Plasmas with High-Frequency Small-Scale Turbulence}

\author{Brett D. Keenan}
\email{bdkeenan@ku.edu}
\affiliation{Department of Physics and Astronomy, University of Kansas, Lawrence, KS 66045}
\author{Mikhail V. Medvedev}
\affiliation{Department of Physics and Astronomy, University of Kansas, Lawrence, KS 66045}


\begin{abstract}
Magnetized high-energy-density plasmas can often have strong electromagnetic fluctuations whose correlation scale is smaller than the electron Larmor radius. Radiation from the electrons in such plasmas, which  markedly differs from both synchrotron and cyclotron radiation, and their energy and pitch-angle diffusion are tightly related. In this paper, we present a comprehensive theoretical and numerical study of the particles' transport in both cold, ``small-scale'' Langmuir and Whistler-mode turbulence and its relation to the spectra of radiation simultaneously produced by these particles. We emphasize that this relation is a superb diagnostic tool of laboratory, astrophysical, interplanetary, and solar plasmas with a mean magnetic field and strong small-scale turbulence.
\end{abstract}

\maketitle

\section{Introduction}
\label{s:intro}

High-energy density plasma environments are generally the sites of turbulent, high-amplitude (i.e., larger or comparable to preexisting ambient magnetic fields) electromagnetic fluctuations, which often exist at scales below a Larmor scales. Such turbulence is a common feature of astrophysical and space plasmas, e.g., at high-Mach-number collisionless shocks and in reconnection regions in weakly magnetized plasmas \citep{medvedev09, frederiksen04, nishikawa03, spitkovsky08, sironi11, sironi13, sironi14}. Additionally, turbulent magnetic fields existing on ``sub-Larmor-scales'' play a critical role in laboratory plasmas; especially in high-intensity laser plasmas, as is observed in experiments at the National Ignition Facility (NIF), OmegaEP, Hercules, Trident, and others \citep{ren04, huntington12, mondal12, tatarakis03}. 

Small-scale electromagnetic turbulence can be of various origin and thus have rather different properties. ``Weibel-like'' turbulence \citep{weibel59, fried59, medvedev09c} may occur in non-magnetized plasmas, i.e.,\  plasmas possessing no ambient (mean) magnetic field. In contrast, several turbulence-producing electromagnetic instabilities require a preexisting magnetic field, e.g.,\ the Whistler-mode, mirror-mode, fire-hose, Bell's-type instability and others \citep{gary06, lazar14, bykov13, bell78, lucek00, bell04, bell05, caprioli14, bai15}.

If the electromagnetic fields are substantially small-scale and statistically random, which is usually the case of turbulence because of the random phases of fluctuations, the paths of the particles diffusively diverge. If the turbulence is sub-Larmor-scale (for the electrons) then the radiation simultaneously produced by the electrons is neither cyclotron nor synchrotron (for non-relativistic or relativistic particles, respectively) but, instead, carries information about the spectrum of turbulent fluctuations. 

In our previous works, we found the relation between the transport of ultrarelativistic \citep{keenan13} and non-relativistic/trans-relativistic \citep{keenan15} particles in isotropic three-dimensional small-scale (mean-free) magnetic turbulence and the radiation spectra simultaneously produced by these particles. In Ref. \citep{keenan13}, we found that the radiation spectrum, in the ultrarelativistic (small deflection angle) regime, agrees with the small-angle jitter radiation prediction \citep{medvedev00,medvedev06,medvedev11,RK10,TT11}. Furthermore, we demonstrated that the pitch-angle diffusion coefficient is directly related to, and can readily be deduced from, the spectra of the emitted radiation. These results were then generalized to non-relativistic and trans-relativistic velocities in Ref. \citep{keenan15}. 

Our previous studies strictly considered a ``Weibel-like'' magnetic turbulence. This means that we treated the electromagnetic turbulence as static, i.e., with zero real frequency and no mean field. In this study, we will extend our model to include sub-Larmor-scale electromagnetic turbulence in plasmas with ambient magnetic fields. The instabilities, in this case, are usually driven with non-zero real frequency, and thus, they induce random electric fields. For this reason, one should not only consider stochastic transport via magnetic pitch-angle diffusion, but transport via electric-field-induced energy and pitch-angle diffusion as well. Additionally, we will show that the energy diffusion coefficient is proportional to the (sub-Larmor-scale) magnetic, pitch-angle diffusion coefficient. The exploitation of the inter-relation between the transport and radiative properties of these plasmas should provide a powerful diagnostic tool for examination of small-scale turbulence in magnetized plasmas.

We will, furthermore, consider the transport of, and radiation production by, relativistic electrons moving through ``small-scale'' Langmuir turbulence -- which is purely electric turbulence. 

Moreover, we omit the resonant wave-particle interactions, which support the underlying electromagnetic turbulence, from our analysis and consider non-resonant particles only -- as the resonant ones constitute a nearly infinitesimal test particle population.

We will, principally, focus on realizations of Whistler-mode turbulence, because Whistler waves are regularly seen in a very wide variety of magnetized enviroments. Given certain conditions, the (temperature anisotropy) Weibel instability -- in pre-magnetized plasmas -- may evolve into a Whistler-mode instability \citep{palodhi10}; thus, for example, Whistler-modes may spontaneously appear in environments where Weibel-like instabilities may take hold.  

Many examples of Whistler waves in space and astrophysical plasmas exist. Whistler waves near collsionless shocks in the solar system, in particular, have been observed {\it in situ} for decades. These wave-modes have, addtionally, been strongly associated with interplanetary shocks -- appearing both in the upstream and downstream regions \citep{fairfield74, tsurutani83, ramirez12}. The solar wind turbulence, as well, appears to host Whistler-modes \citep{lengyel96, lin98}. 

The rest of the paper is organized as follows. Section \ref{s:analytic} presents the analytic theory. Sections \ref{s:model} and \ref{s:results} describe the numerical techniques employed and the obtained simulation results. Section \ref{s:thermal} presents as special case of jitter radiation from a thermal distribution of electrons. Finally, Section \ref{s:concl} is the conclusions. All equations appear in cgs units.

\section{Analytic theory}
\label{s:analytic}

Consider a test particle (electron) moving through a non-uniform, random magnetic field with velocity, ${\bf v}$. Assume that the magnetic field has the mean value, $\langle {\bf B} \rangle$, where $\langle \cdot \rangle$ is an appropriately chosen average over space and, possibly, time. Consequently, we write the total random magnetic field as:
\begin{equation}
{\bf B}({\bf x}, t) = {\bf B}_0 + \delta{\bf B}({\bf x}, t),
\label{mag_def}
\end{equation} 
where ${\bf B}_0 \equiv \langle {\bf B} \rangle$ is the mean field and $\delta{\bf B}({\bf x}, t)$ is the mean-free ``fluctuation'' field, that is $\langle \delta{\bf B} \rangle = 0$ but $\delta B\equiv\langle \delta{B}^2 \rangle^{1/2} \neq 0$. 

Next, the motion of an electron in a random magnetic field is, in general, very complicated. It is the spatial scale of inhomogeneity, i.e., the correlation length of the field fluctuation, that fundamentally determines the dynamics. These magnetic fluctuations are deemed sub-Larmor-scale (or ``small-scale'') when the electron's {\em fluctuation} Larmor radius, $r_L \equiv \gamma\beta m_e c^2/e\delta{B}$ (where $\beta=v/c$ is the dimensionless particle velocity, $m_e$ is the electron mass, $c$ is the speed of light, $e$ is the electric charge, and $\gamma$ is the electron's Lorentz factor) is greater than, or comparable to, the correlation length of the field, $\lambda_B$, i.e., $r_L\gtrsim \lambda_B$. We introduce the ``gyro-number'', which fully characterizes the small-scale regime \citep{keenan15} as follows:
\begin{equation}
\rho \equiv r_L\lambda_B^{-1}.
\label{scale_para}
\end{equation} 
Notice that we are considering only the {\em fluctuation} component of the magnetic field, $\delta{\bf B}$. This is because the motion can be separated into two components: the regular gyro-motion about the mean magnetic field, and the random deflections due to the small-scale random component. In the discussion to follow, we will presuppose that $\rho \gg 1$.

Next, because the {\em fluctuation} Lorentz force on the electron is random, the electron velocity and acceleration vectors vary stochastically, leading to a random (diffusive) trajectory. Additionally, the magnetic Lorentz force acts only upon the component of velocity transverse to the local magnetic field, leading only to energy-conserving (i.e., $\beta = constant$) deflections. Only an electric field can change the particle energy. When this electric field is random, transport via energy diffusion may occur -- we will explore this later.

Ignoring, for the moment, the presence of any electric fields: the electron motion has two limiting regimes -- depending upon the relative strength of the magnetic fluctuations with respect to the mean field. These are a ``straight line'' trajectory with random (transverse) deflections (i.e., $\delta{\bf B} \gg {\bf B}_0$), and a slightly ``perturbed'' helical motion about the mean magnetic field (i.e., $\delta{\bf B} \ll {\bf B}_0$). In the latter case, we will ignore the regular component of the motion. Doing so allows us to consider only the transport in mean-free, small-scale, magnetic turbulence, which we have explored previously. 

This picture is, of course, only correct if any present electric fields are ignored. Electric turbulence, likewise, can induce transport via pitch-angle diffusion -- as we will show later. However, the contribution to the total transport due to electric fields in small-scale Whistler turbulence, specifically, is negligible. 
\newline

\subsection{Transport via Magnetic Pitch-angle Diffusion}
\label{s:pitch_def}

The pitch-angle diffusion coefficient in mean-free, sub-Larmor-scale, magnetic turbulence is a known function of statistical parameters. It may be obtained by considering that the electron's pitch-angle experiences only a slight deflection, $\delta\alpha_B$, over a single magnetic correlation length. Consequently, the ratio of the change in the electron's transverse momentum, $ \Delta p_t$, to its initial momentum, $p$, is $\delta\alpha_B \approx \Delta p_t/p \sim e(\delta B/c)\lambda_B/\gamma m_e v$, since $\Delta p_t \sim F_L\tau_B$ -- where ${\bf F}_L=(e/c)\,{\bf v\times \delta B}$ is the transverse Lorentz force and $\tau_B \sim \lambda_B/v$ is the time to transit $\lambda_B$. The subsequent deflection will be in a random direction, because the field is uncorrelated over the scales greater than $\lambda_B$. As for any diffusive process, the mean squared pitch-angle grows linearly with time. Thus, the diffusion coefficient appears as \citep{keenan13, keenan15}:
\begin{equation}
D_{\alpha\alpha} \equiv \frac{\langle \alpha^2 \rangle}{t} = \frac{\lambda_B}{\gamma^2c\langle \beta_{B}^2 \rangle^{1/2}}{\langle \Omega_{\delta{B}}^2 \rangle},
\label{Daa_def}
\end{equation}
where $\alpha$ is the electron deflection angle (pitch-angle) with respect to the electron's initial direction of motion, $\langle \beta_{B}^2 \rangle^{1/2}$ is an appropriate ensemble-average over the (transverse) electron velocities, and 
\begin{equation}
\boldsymbol\Omega_{\delta{B}} \equiv \frac{e\delta{\bf B}}{m_ec}.
\label{OmegaB_def}
\end{equation} 
In general, the pitch-angle diffusion will be path-dependent, owing to the dependence on the magnetic correlation length, $\lambda_B$. To properly treat the correlation length, we must introduce the two-point autocorrelation tensor of the magnetic fluctuations \citep{keenan15},
\begin{equation}
R^{ij}({\bf r}, t) \equiv \langle \delta{B}^i({\bf x}, \tau)\delta{B}^j({\bf x} + {\bf r}, \tau + t) \rangle_{{\bf x}, \tau},
\label{corr_tensor}
\end{equation}
with the path and time dependent correlation length tensor defined as:
\begin{equation}
\lambda^{ij}_B(\hat{\bf r}, t) \equiv \int_{0}^\infty \! \frac{R^{ij}({\bf r}, t)}{R^{ij}(0, 0)}  \, \mathrm{d}r.
\label{corr_l}
\end{equation}
To evaluate this expression, we must consider the physics involved. In magnetic deflections, only the component of the magnetic field transverse to the particle velocity is involved in the acceleration. Thus, for magnetic fields, we only consider fields transverse to the direction of motion. In contrast, electric fields will have a ``longitudinal'' and ``transverse'' correlation length. The former is important for energy diffusion -- whereas, the latter governs pitch-angle diffusion, since transverse deflections do no work.

Evaluation of Eq.\ (\ref{corr_l}) can be very difficult, in a general case. If we make some simplifying assumptions about the magnetic turbulence, however, we may evaluate Eq. (\ref{corr_l}) exactly. If the transit time of a particle over a correlation length is shorter than the field variability time-scale, then we can treat the magnetic field as static. Additionally, assuming statistical homogeneity and isotropy permits us to use a simple expression for the correlation tensor. 

The pitch-angle diffusion coefficient, under these simplifying assumptions, has been derived previously \citep{keenan15}. We repeat those results here. The magnetic correlation length assumes the form \citep{keenan15}:
\begin{equation}
\lambda_B = \frac{3\pi}{8}\frac{\int_{0}^\infty \! k{|\delta{\bf B}_k|^2}\, \mathrm{d}k}{\int_{0}^\infty \! k^2{|\delta{\bf B}_k|^2}\, \mathrm{d}k},
\label{mag_cor}
\end{equation}
where $|\delta{B}_k|^2$ is the spectral distribution of the fluctuation magnetic field in Fourier ``$k$-space''. Thus, the pitch-angle diffusion coefficient, for sub-Larmor-scale electrons moving through (isotropic/homogeneous) magnetic turbulence, is:
\begin{equation}
D_{\alpha\alpha} = \frac{3\pi}{8}\sqrt{\frac{3}{2}}\frac{\int_{0}^\infty \! k{|\delta{\bf B}_k|^2}\, \mathrm{d}k}{\int_{0}^\infty \! k^2{|\delta{\bf B}_k|^2}\, \mathrm{d}k}\frac{\langle \Omega_{\delta{B}}^2 \rangle}{\gamma^2c\beta},
\label{Daa_specific}
\end{equation}
where we have assumed a mono-energetic distribution of electrons with velocity, $\beta$.

Since the magnetic turbulence is assumed to be statistically homogeneous and isotropic, the deflection angle, $\alpha$, may be chosen with respect to an arbitrary axis. The component of the velocity parallel to ${\bf B}_0$ is unaffected by this mean field. Consequently, without loss of generality, we can define $\alpha$ as the conventional pitch-angle -- i.e., the angle of the velocity vector with respect to the mean (ambient) magnetic field.

\subsection{Pitch-angle Diffusion in Small-Scale Electric Fields}
\label{s:elec_pitch}

The derivation for pitch-angle diffusion in general small-scale electric turbulence follows in a similar fashion. Suppose an electron test particle is moving, with speed $v$, through an external random electric field. This may be an electrostatic field (i.e., Langmuir-like turbulence), or -- as in the more general case -- it may be the electric component of electromagnetic turbulence (e.g. Whistler-mode turbulence). We will assume that the electric field fluctuates very slowly -- such that the particle dynamics, on relevant time-scales, are largely unaffected by the field's time-variability. Furthermore, we will ignore any present magnetic fields -- for the moment.

For ``small-scale'' turbulence, the principal time-scale which governs particle transport is the time to transit a single electric field correlation length, $\lambda_E^t$ -- where the ``$t$'' superscript indicates that the correlation length is specified along the path with a ``transverse'' component of the electric field (which we did with the magnetic field). If the (pitch-angle) transit time, $\tau_E^t \sim \lambda_E^t/v$, is much less than the field-variability time-scale, $\Omega^{-1}$, then we may treat the electric field as approximately time-independent. 

To proceed, it will be instructive to first discuss the radiation produced by an electron moving through an external random field. First, regardless of the acceleration mechanism, the radiation of an ultrarelativistic electron will be beamed along a narrow cone with opening angle, $\Delta{\theta} \sim 1/\gamma$. In a random electromagnetic field, the acceleration occurs principally along the extent of a correlation length. Since the electron is moving ultrarelativistically, it will undergo a slight deflection, $\delta\alpha_E$, as it traverses a correlation length. If $\delta\alpha_E \ll \Delta{\theta}$, then the electron will move approximately rectilinearly, undergoing only slight random deflections along its path; the radiation will then be beamed along the extent of the electron's relatively fixed direction of motion. Consequently, an observer on axis would see a signal for the entire trajectory of the electron. Furthermore, the radiation spectrum will be wholly determined by the statistical properties of the underlying acceleration mechanism \citep{landau75}. When the acceleration mechanism is a random (static) magnetic field, the electron emits radiation in the small-angle jitter regime \citep{medvedev00, medvedev06, medvedev11, RK10, TT11, keenan13, keenan15}. The radiation produced by ultrarelativistic electrons moving through electrostatic turbulence, in this small deflection angle regime, is nearly identical -- which has lead to its designation as a subclass of small-angle jitter radiation \citep{teraki14}.

We have previously shown that these random deflections initiate pitch-angle diffusion in sub-Larmor-scale magnetic turbulence, and that this diffusion coefficient is intimately related to the radiation spectrum \citep{keenan13, keenan15}. We expect that an electric field analog of this diffusion exists for the (small-angle) jitter regime in small-scale electric turbulence. Here, we consider an electric field as ``small-scale'', with respect to the test electrons, if: 
\begin{subequations}     
\begin{align}
\Omega^{-1} \gg \tau_E^t,  \\  
 \Delta{\theta} \gg \delta\alpha_E.
\end{align}
\label{small_scale_def}
\end{subequations}
Since the electron is moving ultrarelativistically, the component of its acceleration transverse to its direction of motion will be far larger than the longitudinal component. Thus, its motion occupies the small deflection angle regime -- which is the reason its radiation spectrum resembles the jitter spectrum. Additionally, transverse accelerations leave the particle's kinetic energy fixed. For this reason, we will assume a constant $v$. 

Next, since the deflections are small, $\delta\alpha_E \sim \Delta{p}_t/p$ -- as previously noted for magnetic deflections. Since $\Delta{p}_t/\tau_E^t \sim eE_t$, where $E_t$ is the component of the electric field perpendicular (transverse) to the electron's direction of motion, $\Delta{p}_t/p \sim eE_t/\gamma{m_e}v$; thus:
\begin{equation}
\delta\alpha_E \sim \frac{eE_t}{\gamma{m_e}v}\tau_E^t.
\label{alph_def}
\end{equation}
Consequently, the electric diffusion coefficient must be:
\begin{equation}
D_{\alpha\alpha}^{\text{elec.}} \sim \delta\alpha_E^2/\tau_E^t \sim \frac{e^2E_t^2}{\gamma^2m_e^2v^2}\left(\frac{\lambda_E^t}{v}\right).
\label{daa_cor}
\end{equation}
Finally, the exact numerical coefficients depend upon the statistical properties of the turbulent fluctuations. Given statistically isotropic and homogeneous turbulence, $\langle E_t^2 \rangle = \frac{2}{3}\langle E^2 \rangle$. Thus, the diffusion coefficient follows as:
\begin{equation}
D_{\alpha\alpha}^{\text{elec.}}  = \frac{2}{3}\frac{\lambda_E^t}{\gamma^2c\beta^3}\langle \Omega_E^2 \rangle,
\label{daa_elec_def}
\end{equation}
where:
\begin{equation}
\Omega_E \equiv eE/m_ec.
\label{oe_def}
\end{equation}
When a magnetic field is introduced, the (small-scale) pitch-angle diffusion coefficient will be the sum of the magnetic and electric components -- i.e.\ Eq.\ (\ref{daa_elec_def}) and Eq.\ (\ref{Daa_def}).

As we mentioned previously, and will demonstrate later, the electric pitch-angle diffusion is negligibly small compared to the magnetic equivalent in small-scale Whistler-mode turbulence. For this reason, the electric contribution to the radiation production is, also, insignificant. Nonetheless, the electric field will still uniquely affect the particle motion via energy diffusion.

\subsection{Energy Diffusion in Small-Scale Electric Turbulence}
\label{s:energy_def}

All electromagnetic turbulence results from instabilities, dynamo-action, etc.\ with some finite growth rate. So long as the growth (or dissipation) time-scale is much greater than the correlation length transit time, we can ignore the time-dependence of the magnetic field in our model.

In contrast to Weibel magnetic fields in (initially) unmagnetized plasmas, however, MHD/kinetic instabilities (which require an ambient magnetic field) may grow random fields with non-negligible real frequency, $\Omega_r$. That is to say, these magnetic fields will possess oscillating wave-modes, whose time-dependence may not be completely ignored. The Faraday-induced electric fields, ${\bf E}$, may influence the particle motion on relevant time-scales, e.g., the gyro-period time-scale in the regular (ambient) magnetic field.

These random electric fields may induce transport via energy diffusion. Although diffusive energy transport in electromagnetic turbulence has long been a topic of investigation \citep{stix92}, energy diffusion in strictly sub-Larmor-scale electromagnetic fields has yet to be -- to the best of our knowledge -- explored. This topic has proved to be richly complicated, so we have limited ourselves to a particularly simple regime. 

Furthermore, we emphasis that the ``energy'' diffusion coefficient -- rather than the ``velocity-space'' analog -- is a more useful quantity for our purposes. Although it possesses a number of favorable properties, its prominent feature is that it is directly proportional to the electric field's correlation length. This feature is not present in the ``velocity-space'' coefficient, however.  

Next, we must consider the time-scales involved. There are two such characteristic time-scales: the ``acceleration'' time, $\tau_E^l$ and the electric field ``auto-correlation'' time, $\tau_{ac}$. The latter time-scale characterizes the temporal inhomogeneity of the electric field. Diffusive (energy) transport may arise not only from spatial stochasticity in the electric field but temporal randomness as well.

The former quantity, $\tau_E^l$, characterizes the spatial stochasticity. This is the time required to transit an electric field correlation length, $\lambda_E^l$ -- with the ``$l$'' superscript indicating the ``longitudinal'' transit time; i.e.\ the time required to transverse a ``longitudinal'' electric correlation length, $\lambda_E^l$, which is along the direction of motion. Assuming that $a_\lambda{\tau_E^l} \ll v_E$, where $a_\lambda$ is the acceleration over $\lambda_E^l$, and $v_E$ is the component of the electron velocity parallel to the electric field, the transit time is:
\begin{equation}
\tau_E^l \sim \frac{\lambda_E^l}{v_E}.
\label{tau_elec}
\end{equation}
While transiting a single correlation length, the electron is subject to a nearly uniform electric field. These ``accelerations'' are uncorrelated on a spatial-scale dictated by the electric field correlation length. 

The diffusion regime we will explore will consider the ``spatial'' diffusion to be the dominant process, i.e.,
\begin{equation}
\tau_E^l \ll \tau_{ac}.
\label{energy_regime0}
\end{equation}
Furthermore, to ensure that the energy change is random on the time-scale of consideration, we require that: 
\begin{equation}
\tau_E^l \ll t.
\label{energy_regime1}
\end{equation}
Next, an equation for the electron energy, $W_e$, may be obtained directly from the Lorentz Force Equation of Motion. It is:
\begin{equation}
\frac{dW_e}{dt} = e\left({\bf v}\cdot{\bf E}\right).
\label{energy_eom}
\end{equation}
Since the electron energy changes over the characteristic time-scale, $\tau_E^l$, we may write:
\begin{equation}
\frac{\Delta{W}_\lambda}{\tau_E^l} \sim e{v}_{E}E.
\label{energy_change_def}
\end{equation}
If the random process is, indeed, diffusive:
\begin{equation}
 D_{WW} \equiv \frac{\langle W_e^2 \rangle}{t}.
\label{Energy_diff_def}
\end{equation}
Thus:
\begin{equation}
D_{WW} \sim \frac{\left(\Delta{W}_\lambda\right)^2}{\tau_E^l} \sim e^2v_EE^2\lambda_E^l,
\label{energy_diff}
\end{equation}
where we have used Eq. (\ref{tau_elec}). With the usual assumptions of statistical homogeneity/isotropy and an initially mono-energetic distribution of electrons, we may write the energy diffusion coefficient, thusly:
\begin{equation}
D_{WW} = \sqrt{\frac{1}{3}}e^2{\langle{E^2}\rangle}v\lambda_E^l.
\label{energy_diff_def}
\end{equation}

This result may be contrasted with the ``temporal'', i.e.\ {\it resonant}, energy diffusion coefficient. The physics of this type of diffusion may be understood by considering the, so called, ``Quasilinear'' energy diffusion coefficient. As before, we will consider only small corrections to the electron's initial velocity -- hence, we will assume the zero-order trajectory:
\begin{equation}
{\bf r}(t) = {\bf v}t + {\bf r}_0,
\label{elec_traj}
\end{equation}
where ${\bf r}_0$ is the electron's initial position. Let us suppose that the electric field assumes a simple sinusoidal profile, i.e.\ 
\begin{equation}
{\bf E}({\bf x}, t) = {\bf E}_0\text{cos}({\bf k}\cdot{\bf x} - \Omega{t}).
\label{elec_traj}
\end{equation}
Thus, using Eqs.\ (\ref{energy_eom}) and (\ref{elec_traj}), we have:
\begin{equation}
\frac{dW_e}{dt} = e\left({\bf v}\cdot{\bf E}_0\right)\text{cos}({\bf k}\cdot{\bf v}t +{\bf k}\cdot{\bf r}_0  - \Omega{t}).
\label{dwdt_res}
\end{equation}
Integrating Eq.\ (\ref{dwdt_res}), averaging over all possible initial positions, and squaring the result, gives the energy variance:
\begin{equation}
\langle \Delta{W_e}^2 \rangle = \left[\frac{e\left({\bf v}\cdot{\bf E}_0\right)}{\left(\Omega - {\bf k}\cdot{\bf v}\right)}\right]^2\text{sin}^2\left[\frac{\left(\Omega - {\bf k}\cdot{\bf v}\right)t}{2}\right].
\label{dw2_avg}
\end{equation}
Finally, with $\Omega{t} \gg 1$, we may employ the relation \citep{stix92}:
\begin{equation}
\text{sin}^2\left[\frac{\left(\Omega - {\bf k}\cdot{\bf v}\right)t}{2}\right] \sim \pi\delta\left(\Omega - {\bf k}\cdot{\bf v}\right),
\label{sine_approx}
\end{equation}
Thus the (Quasilinear) diffusion coefficient is:
\begin{equation}
D^\text{res.}_{WW} \equiv \frac{\langle \Delta{W_e}^2 \rangle}{t} \sim \pi\left[\frac{e\left({\bf v}\cdot{\bf E}_0\right)}{\left(\Omega - {\bf k}\cdot{\bf v}\right)}\right]^2\delta\left(\Omega - {\bf k}\cdot{\bf v}\right).
\label{dw2_avg}
\end{equation}
In general, turbulence will contain a spectrum of waves; hence, an integration of Eq.\ (\ref{dw2_avg}) over $|{\bf E}_{{\bf k}, \Omega}|^2$ is required to produce the complete diffusion equation. 

Nevertheless, much can be gathered by examining the functional form of this simplified expression. For example, owing to the dependence on the quantity, $\delta\left(\Omega - {\bf k}\cdot{\bf v}\right)$, only particles that are in resonance with the wave participate in the diffusive process.

 Moreover, since $\Omega{t} \gg 1$, this ``temporal'' diffusion process occurs on a much greater time-scale than $\tau_E^l$ (when the electric field is small-scale). For this reason, the non-resonant energy diffusion coefficient -- Eq.\ (\ref{energy_diff}) -- is much greater than the resonant equivalent -- at least, for the ``small-scale'' population of electrons.

As an important side note, the``Quasilinear'' diffusion equation derived here applies for non-magnetized plasmas. When an ambient magnetic field, ${\bf B}_0$, is present, the ``resonance'' condition generalizes to \citep{stix92}: 
\begin{equation}
\Omega - k_{\parallel}v_{\parallel} = n\Omega_\text{ce}/\gamma,
\label{res_mag}
\end{equation}
where $\Omega_{ce} \equiv eB_0/m_ec$ is the non-relativistic gyro-frequency, the ``parallel'' direction is along the ambient (mean) magnetic field, and $n$ is an integer. Electrons moving through electromagnetic turbulence are not ``magnetized'', in the formal sense, with respect to the ``small-scale'' fluctuation fields. Hence, the small-scale fields do not contribute to the higher-order (magnetic) resonances -- such as the Cherenkov resonance at $n = 1$. Thus, with regard to the ``small-scale'' sub-population of electrons, we may disregard resonant diffusion in general.

Finally, to evaluate the (non-resonant) energy diffusion coefficient -- Eq.\ (\ref{energy_diff_def}) -- we need an expression for the electric field, $\langle E^2 \rangle$, and its ``longitudinal'' correlation length, $\lambda_E^l$. To this end, we must relate the electric field to the underlying magnetic turbulence that produces it, i.e., we need to specify the wave turbulence dispersion relation, $\Omega_r({\bf k})$. 

In general, this may be done via the dielectric tensor, $\stackrel{\leftrightarrow}{\epsilon}_{{\bf k}, \Omega}$. Using Amp$\grave{\text{e}}$re's law, and the definition of the dielectric tensor, we write \citep{brambilla}:
\begin{equation}
{\bf k}\times\delta{\bf B}_{{\bf k}, \Omega} = -\frac{\Omega}{c}\stackrel{\leftrightarrow}{\epsilon}_{{\bf k}, \Omega}\cdot \ {\bf E}_{{\bf k}, \Omega}.
\label{eb_rel}
\end{equation}
Suppressing the time-dependent in the field amplitudes, i.e.\ ignoring wave growth/damping, the electric spectral distribution may be expressed as:
\begin{equation}
\left|{\bf E}_{\bf k}\right|^2 = \left|\stackrel{\leftrightarrow}{\epsilon}_{{\bf k}, \Omega}^{-1}\cdot \ \hat{b}_{\bf k}^t\right|^2n^2\left|\delta{\bf B}_{\bf k}\right|^2,
\label{eb_spec}
\end{equation}
where $\hat{b}_{\bf k}^t$ is the unit vector in the direction of ${\bf k}\times\delta{\bf B}_{{\bf k}, \Omega}$, and $n \equiv kc/\Omega$ is the index of refraction. 

Next, using Eq.\ (\ref{eb_spec}) and Parseval's theorem, $\langle E^2 \rangle$  becomes:
\begin{equation}
\langle E^2 \rangle = \frac{\int \! \left|\stackrel{\leftrightarrow}{\epsilon}_{{\bf k}, \Omega}^{-1}\cdot \ \hat{b}_{\bf k}^t\right|^2n^2{\left|\delta{\bf B}_{\bf k}\right|^2}\, \mathrm{d}{\bf k}}{\int \! {\left|\delta{\bf B}_{\bf k}\right|^2}\, \mathrm{d}{\bf k}}\langle \delta{B}^2 \rangle.
\label{eb_rms}
\end{equation}
Finally, the general expression which relates the (electric) energy diffusion and (magnetic) pitch-angle diffusion coefficients follows from Eqs.\ (\ref{eb_rms}), (\ref{energy_diff_def}), and (\ref{Daa_def}). It is:
\begin{equation}
D_{WW} = \frac{\sqrt{2}}{3}W_e^2\beta^2\frac{\int \! \left|\stackrel{\leftrightarrow}{\epsilon}_{{\bf k}, \Omega}^{-1}\cdot \ \hat{b}_{\bf k}^t\right|^2n^2{\left|\delta{\bf B}_{\bf k}\right|^2}\, \mathrm{d}{\bf k}}{\int \! {\left|\delta{\bf B}_{\bf k}\right|^2}\, \mathrm{d}{\bf k}}\left(\frac{\lambda_E^l}{\lambda_B}\right)D_{\alpha\alpha},
\label{dww_daa}
\end{equation}
where $W_e \equiv \gamma m_ec^2$ is the electron's total energy, and we have assumed statistical isotropy/homogeneity to produce the numerical prefactor. 

Eq.\ (\ref{dww_daa}), despite its apparent complication, offers a fairly simple interpretation when the dielectric tensor assumes a scalar value, $\epsilon$. Recalling that $\sqrt{\epsilon} = n$, so that $\epsilon^{-1} = 1/n^2$, Eq.\ (\ref{dww_daa}) simplifies to:
\begin{equation}
D_{WW} = \frac{\sqrt{2}}{3}W_e^2\beta^2\langle \beta_{ph}^2 \rangle_\text{dist.} \left(\frac{\lambda_E^l}{\lambda_B}\right)D_{\alpha\alpha},
\label{dww_daa_scalar}
\end{equation}
where $\langle \beta_{ph}^2 \rangle_\text{dist.}$ is the distributional average, over the magnetic spectrum, of the normalized wave phase velocity, $\beta_{ph} \equiv \Omega/kc$. Thus, 
\begin{equation}
D_{WW} \propto \left(m_e^2v^2\langle v_{ph}^2 \rangle_\text{dist.}\right)D_{\alpha\alpha},
\label{dww_daa_propto}
\end{equation}
which is what we would expect, given the general relation between the ``velocity space'' diffusion coefficient, $D_{vv}$, and the pitch-angle diffusion coefficient; i.e.\ $D_{vv} \sim v_{ph}^2D_{\alpha\alpha}$ \citep{cravens}.

\subsection{Particle Transport in Magnetized Plasmas with Electric Fluctuations}
\label{s:par_trans_elec}

As mentioned previously, the combined effect of electric and magnetic fields can lead to fairly complicated particle dynamics. Particle {\it drifts}, for example, involving both the electric and magnetic fields, should be considered. Here, we present two realizations of the drift phenomenon. We will, subsequently, argue that these effects are of negligible importance for diffusion in small-scale fields.

In Section \ref{s:pitch_def}, we argued that sub-Larmor-scale magnetic fluctuations result in trajectories that occupy the small deflection angle regime. For this reason, the ``guiding center approximation'', that underlies the drift theory, breaks down. Consequently, the notions of {\it curvature drift} and {\it Grad-B drift} lose all meaning in this regime.

Nonetheless, a magnetized plasma contains a large-scale magnetic field -- which is, by construction, ``super-Larmor-scale''. Hence, drifts that involve the electric field and the ambient (mean) field are, in principle, important to consider.

The first of these that we will explore is the, so called, {\it E cross B drift}. We will, once more, assume a sinusoidal electric field. In this case, however, we assume that an ambient magnetic field, ${\bf B}_0$, is present. Furthermore, we suppress the time-dependence; hence:
\begin{equation}
{\bf E}(x) = E_0\text{cos}(kx)\hat{x},
\label{dww_daa_scalar}
\end{equation}
where the $x$-direction is along ${\bf k}$. Assuming non-relativistic velocities, the $y$-component of the electron, in the ambient magnetic field, will have the solution \citep{chen}:
\begin{equation}
\frac{d^2v_y}{dt^2} = -\Omega_\text{ce}^2v_y - \Omega_\text{ce}^2\frac{cE_0}{B_0}\text{cos}\left[kx_0 + kr_{L0}\text{sin}(\Omega_\text{ce}t)\right]
\label{eb_eom}
\end{equation}
where $x_0$ is the initial position, and $r_{L0} = {m_e}\beta{c^2}/eB_0$ is the (ambient) Larmor radius. This solution presupposes that the electric field will only perturb the electron orbit about the ambient field. Hence, our substitution of the zero$^{th}$-order solution.

Next, we average Eq.\ (\ref{eb_eom}) over a gyro-period. Thus,
\begin{equation}
\langle v_y \rangle + \frac{cE_0}{B_0}\langle \text{cos}\left[kx_0 + kr_{L0}\text{sin}(\Omega_\text{ce}t)\right]\rangle = 0,
\label{eb_eom_avg}
\end{equation}
since $\langle dv_y/dt \rangle = 0$ -- i.e.\ the drift velocity is constant.

Next, assuming that $kr_{L0} \ll 1$, we may write the solution for $\langle v_y \rangle$ as \citep{chen}:
\begin{equation}
\langle v_y \rangle/c = -\frac{E(x)}{B_0}\left(1 - \frac{1}{4}k^2r_{L0}^2\right).
\label{eb_k_sol}
\end{equation}
Finally, recognizing that, in the general case, $i{\bf k} \rightarrow {\bf \nabla}$, we write the solution for an arbitrary electric field as \citep{chen}:
\begin{equation}
{\bf v}_{{\bf E}\times{\bf B}} = c\left(1 + \frac{1}{4}r_{L0}^2\nabla^2\right)\frac{{\bf E}\times{\bf B}_0}{B_0^2},
\label{eb_sol}
\end{equation}
where ${\bf v}_{{\bf E}\times{\bf B}}$ is the drift velocity. The second term, i.e.\ that which involves the Laplacian operator, is a correction known as a {\it finite-Larmor-radius effect}. When $kr_{L0} \gg 1$, the Larmor radius is much larger than the field wavelength. In this case, the particle is acted upon, by the electric field, on a time-scale much shorter than the gyroperiod. Consequently, the drift approximation is not appropriate for ``small-scale'' fields, since the perturbation is implicitly assumed to act on a time-scale of many gyroperiods.

A similar drift phenomenon occurs when we consider the time-dependence of the electric field. Assuming that $\Omega^2 \ll \Omega^2_\text{ce}$, the particle will drift with velocity \citep{chen}:
\begin{equation}
{\bf v}_p = \pm\frac{c}{\Omega_\text{ce}B_0}\frac{d{\bf E}}{dt},
\label{dedt_sol}
\end{equation}
The quantity, ${\bf v}_p$, is known as the {\it polarization drift} velocity. Similarly, the small-scale processes -- by construction -- occur on time-scales much shorter than $\Omega^{-1}$. Hence, the polarization drift time-scale will be far greater than either $\tau_E^l$ or $\tau_E^t$. For this reason, polarization drift is not significant on the time-scales of immediate interest.

In the next subsection, we will consider the case of small-scale energy diffusion in isotropic, small-scale Whistler turbulence.

\subsection{Energy Diffusion in Small-Scale Whistler Turbulence}
\label{s:energy_whist_diff}

Next, to evaluate Eq. (\ref{energy_diff_def}), we consider a concrete example of electromagnetic turbulence in a magnetized plasma. Whistler-mode turbulence in a ``cold'' plasma admits a simple dispersion relation \citep{sazhin93}:
\begin{equation}
\Omega_r(k) = \Omega_{ce}\frac{k^2c^2}{k^2c^2 + \omega_{pe}^2}\cos(\theta_k),
\label{whis_disp}
\end{equation} 
where $\omega_{pe} \gg \Omega_{ce}$ is the electron plasma frequency, and $\theta_k\in (0, \pi/2)$, is the angle between the wave-vector, ${\bf k}$, and the ambient magnetic field, ${\bf B}_0$. We will assume a (nearly) steady-state, so that the instability is non-linearly saturated, that is the instability growth rate, $\Omega_i$, is much less than all relevant frequency-scales, and thus is negligible. This treatment assumes that the turbulence is ``linear'', i.e.\ $\delta{B} \ll B_0$. We will further assume that:
\begin{equation}
\frac{\gamma v}{\Omega_{ce}} > \lambda_B,
\label{mean_condition}
\end{equation} 
where $\Omega_{ce}/\gamma$ is the relativistic gyro-frequency.

Eq. (\ref{mean_condition}) implies that $\rho \gg 1$, since $\delta{B} \ll B_0$ -- thus, the test electrons are sub-Larmor-scale with respect to the fluctuation magnetic field, ${\bf \delta{B}}$.

It is worth mentioning that, formally, the cold plasma approximation requires that $kv/\Omega_{ce} \ll 1$ \citep{verkhoglyadova10}. This condition would imply that the electron population is ``super-Larmor-scale'' with respect to the magnetic field, since $\lambda_B \sim k_B^{-1}$, where $k_B$ is the wave-number of the dominant wave-mode. For this reason, our model implicitly presupposes the existence of a cold population of super-Larmor-scale electrons which support the Whistler-modes. Consequently, our test particles will be comprised of a ``hot'', albeit smaller, population of sub-Larmor-scale electrons. This situation may be approximately realized by the ``Super-halo'' electron population \citep{lin98b, wang15}, as it propagates through the ``colder''  solar wind turbulence, which appears to include small-scale Kinetic-Alfv{\' e}n and Whistler-wave modes \citep{che14}.

An examination of Eq. (\ref{whis_disp}) reveals that $\Omega_r(k) \ll \Omega_{ce}$ in the regime where $kc \ll \omega_{pe}$. Restricting ourselves to this regime motivates the introduction of a new parameter, which we call the ``skin-number''. It is:
\begin{equation}
\chi \equiv d_e\lambda_B^{-1},
\label{skin_num_def}
\end{equation} 
where $d_e \equiv c/\omega_{pe}$, is the electron skin-depth. Thus, the regime of interest is characterized by $\chi \ll 1$. 

It is noteworthy that, in principle, the test electron velocities may be large enough so that $\Omega_{ce}/\gamma \sim \Omega_r$. By restricting the electron velocities to the mildly relativistic regime, we may safely presuppose that the field-variability time, $\Omega_r^{-1}$, is sufficiently greater than the time to transit a magnetic correlation length, thus permitting the static field treatment for the magnetic field and avoiding the wave-particle resonance treatment.

Next, in the $\chi \ll 1$ regime, the electric field perpendicular to ${\bf B}_0$ is much greater than the component parallel to the ambient magnetic field; i.e.\ $E_{\perp} \gg E_{\parallel}$ \citep{sazhin93}. Furthermore, it can be shown that in the frame moving along the direction of ${\bf B}_0$ with velocity equal to the parallel phase velocity, $v^{\parallel}_{ph} \equiv \Omega_r/k_{\parallel}$, the perpendicular electric field is approximately zero \citep{sazhin93}. Consequently, this allows us, via Lorentz transformation of the electromagnetic fields, to relate the magnetic spectral distribution to the electric distribution. It is, thusly: 
\begin{equation}
|{\bf E}_{\bf k}|^2 \approx |{\bf E}^{\perp}_{\bf k}|^2 \approx \beta_{ph}^2|\delta{\bf B}^{\perp}_{\bf k}|^2, 
\label{eq:D3}
\end{equation}
where $\perp$ refers to the spectrum perpendicular to the mean magnetic field, and
\begin{equation}
\beta_{ph} \equiv \frac{v^{\parallel}_{ph}}{c} = \frac{\Omega_r({\bf k})}{k_\parallel{c}}.
\label{beta_ph_def}
\end{equation}
Given isotropic/homogeneous magnetic turbulence:
\begin{equation}
|\delta{\bf B}^{\perp}_{\bf k}|^2 = |\delta{\bf B}_{k}|^2\cos^2(\theta_k). 
\label{eq:D4}
\end{equation}
This relation then allows us to express $\langle E^2 \rangle$ in terms of the magnetic field as:
\begin{equation}
\langle E^2 \rangle = \frac{2}{3}\frac{\int \!  \beta_{ph}^2 {|\delta{\bf B}_{k}|^2}\, \mathrm{d}{\bf k}}{\int \! {|\delta{\bf B}_{k}|^2}\, \mathrm{d}{\bf k}}\langle \delta{B}^2 \rangle.
\label{E2_def}
\end{equation}
Next, the electric field correlation length may be obtained from the electric field correlation tensor. For isotropic turbulence, one may write the general expression for the Fourier image of the electric field two-point auto-correlation tensor as:
\begin{equation}
\Phi_{ij}({\bf k}) = |{\bf E}^{t}_{\bf k}|^2(\delta_{ij} - \frac{k_ik_j}{k^2}) + |{\bf E}^{l}_{\bf k}|^2\frac{k_ik_j}{k^2}.
\label{elec_cor_tens}
\end{equation}
Isotropy is an approximation here, given the polar asymmetry indicated by Eq. (\ref{eq:D4}). Using Maxwell's Equations, we may relate the longitudinal, $|{\bf E}^{l}_{\bf k}|^2$ and transverse, $|{\bf E}^{t}_{\bf k}|^2$ distributions to $|\delta{\bf B}_{k}|^2$ (where ``longitudinal'' and ``transverse'' are with respect to the wave-vector, not the electron velocity). To wit:
\begin{equation}
\left\{\begin{array}{ll}
|{\bf E}^{t}_{\bf k}|^2 = \frac{\Omega_r^2}{k^2c^2}|\delta{\bf B}_{k}|^2 \\
|{\bf E}^{l}_{\bf k}|^2 = |{\bf E}_{\bf k}|^2 - |{\bf E}^{t}_{\bf k}|^2  \\
\end{array}\right.
\label{elec_bs}
\end{equation}
In the $\chi \ll 1$ regime, we may substitute Eq. (\ref{eq:D3}) to express the tensor completely in terms of the magnetic spectrum. The trace of the correlation tensor is then given by:
\begin{equation}
Tr\left[\stackrel{\leftrightarrow}{\Phi}({\bf k})\right] = 2 \beta_{ph}^2|\delta{\bf B}_{k}|^2\cos^2(\theta_k).
\label{ecor_trace}
\end{equation}
While integrating Eq. (\ref{ecor_trace}) along a selected path, we only consider the component of the electric field parallel to the trajectory, owing to the dot product with velocity in Eq. (\ref{energy_eom}). This allows us to draw an analogy to the ``monopolar'' (magnetic) correlation length considered in Ref. \citep{keenan15} -- permitting us to write the expression immediately as:
\begin{equation}
\lambda_E^l \equiv \lambda^{Tr}_E(x\hat{x}) = \frac{3\pi}{4}\frac{\int \! (v^{\parallel}_{ph})^2k{|\delta{\bf B}_{k}|^2}\, \mathrm{d}{k}}{\int \! (v^{\parallel}_{ph})^2k^2{|\delta{\bf B}_{k}|^2}\, \mathrm{d}{k}},
\label{cor_E}
\end{equation}
where the integration path was chosen to be along the x-axis. By comparing Eq. (\ref{mag_cor}) to Eq. (\ref{cor_E}), we see that the electric correlation length differs from the magnetic correlation length only by a factor of a few. For this reason, we may conclude that $\tau_E^l$ is less than $\tau_{ac} \sim \Omega_r^{-1}$. Consequently, the energy diffusion will be dominated by the electric field's ``spatial'' stochasticity, as per Eq. (\ref{energy_regime0}). 

Additionally, $\chi \ll 1$ and $\Omega_{ce} \ll \omega_{pe}$ demand that $v^{\parallel}_{ph} \ll c$. This implies that $\langle \delta{B}^2 \rangle \gg \langle E^2 \rangle$. Consequently, the pitch-angle diffusion will be dominated by the magnetic deflections, and thus we may neglect the contribution due to the electric field. 

Finally, given Eq. (\ref{E2_def}), the energy diffusion coefficient may be related directly to the (magnetic) pitch-angle diffusion coefficient via the relation:
\begin{equation}
D_{WW} = \frac{2\sqrt{2}}{9}W_e^2\beta^2\frac{\int \! (\beta^{\parallel}_{ph})^2{|\delta{\bf B}_{k}|^2}\, \mathrm{d}{\bf k}}{\int \! {|\delta{\bf B}_{k}|^2}\, \mathrm{d}{\bf k}}\frac{\lambda_E^l}{\lambda_B}D_{\alpha\alpha}.
\label{Dww_rel_daa}
\end{equation}
Eqs. (\ref{Dww_rel_daa}) and (\ref{Daa_specific}) will be confirmed, given isotropic small-scale Whistler turbulence, via first-principle numerical simulation in Section \ref{s:results}.

\subsection{Radiation Production in Magnetized Plasmas with Sub-Larmor-Scale Magnetic Fluctuations}
\label{s:rad_def}

As mentioned previously, radiation production by electrons moving through (mean-free) sub-Larmor-scale magnetic turbulence has been explored thoroughly by a number of authors. The ultrarelativistic regime, specifically, is characterized by a single parameter, the ratio of the deflection angle, $\delta\alpha_B$ (over a single magnetic correlation length) to the relativistic beaming angle, $\Delta\theta \sim 1/\gamma$. The ratio \citep{medvedev00, medvedev11, keenan13}:
\begin{equation}
\frac{\delta\alpha_B}{\Delta\theta} \sim \frac{e\delta{B}}{m_e c^2}\lambda_B \equiv \delta_j,
\label{delta}
\end{equation} 
is known as the jitter parameter. If $\delta_j \ll 1$, which implies that $\rho \gg 1$, then a distant observer on the line-of-sight will see the radiation along, virtually, the entire trajectory of the particle. This radiation is known as small-angle jitter radiation \citep{medvedev00, medvedev06, medvedev11}. Jitter radiation is distinctly not synchrotron radiation. The jitter radiation spectrum is wholly determined by $\delta_j$ and the magnetic spectral distribution. Consider an isotropic power-law magnetic spectrum for a time-independent field, such as:
\begin{equation}
\left\{\begin{array}{ll}
|{\bf B}_{\bf k}|^2 = Ck^{-\mu}, & k_{min} \le k \le k_{max} 
 \\
|{\bf B}_{\bf k}|^2 = 0, & \text{otherwise}
\end{array}\right.
\label{Bk}
\end{equation} 
where the magnetic spectral index, $\mu$ is a real number, and $C$ is a normalization. It has been shown \citep{medvedev06, medvedev11,RK10,TT11} that monoenergetic ultrarelativistic electrons in this prescribed sub-Larmor-scale turbulence produce a flat angle-averaged spectrum below the spectral break and a power-law spectrum above the break, that is:
\begin{equation}
P(\omega) \propto 
\left\{\begin{array}{ll}
\omega^0, &\text{if}~ \omega<\omega_j, \\
\omega^{-\mu + 2}, &\text{if}~ \omega_j<\omega<\omega_b, \\
0, &\text{if}~ \omega_b<\omega,
\end{array}\right.
\label{Pomega}
\end{equation}
where the spectral break is
\begin{equation}
\omega_j =\gamma^2 k_\textrm{min} c, 
\label{omegaj-kmin}
\end{equation}
which is called the jitter frequency. Similarly, the high-frequency break is 
\begin{equation}
\omega_b =\gamma^2 k_\textrm{max} c.
\label{omegab}
\end{equation}
Recently, we have generalized the small-scale jitter regime to non-relativistic and mildly relativistic velocities \citep{keenan15}. Non-relativistic jitter radiation, or ``pseudo-cyclotron'' radiation, differs markedly from both synchrotron and cyclotron radiation. Since relativistic beaming is not realized in the non-relativistic regime, the jitter parameter loses its meaning here. Instead, the gyro-number characterizes the regime, i.e.\ $\rho \gg 1$. Given a monoenergetic distribution of electrons, and the spectral distribution indicated by Eq. (\ref{Bk}), the pseudo-cyclotron spectrum has a slightly more complicated structure than ultrarelativistic jitter radiation. It appears as \citep{keenan15}:
\begin{equation}
P(\omega) \propto 
\left\{\begin{array}{ll}
A + D\omega^2, &\text{if}~ \omega \leq \omega_\textrm{jn}
 \\
F\omega^{-\mu+2} + G\omega^2 + K, &\text{if}~ \leq \omega_\textrm{bn} 
 \\
0, &\text{if}~  \omega > \omega_\textrm{bn},
\end{array}\right.
\label{analy_spec}  
\end{equation} 
where $\mu \neq 2$ and $A$, $D$, $F$, $G$, and $K$ are functions of spectral/particle parameters (e.g.\ $\mu$, $k_\textrm{min}$, and $\beta$). The break frequencies generalize to non-relativistic velocities as expected; namely:
\begin{equation}
\omega_\textrm{jn} = k_\textrm{min}\beta{c},
\label{omega_jn_nonrel}  
\end{equation} 
\newline
and the break frequency indicated by the smallest spatial scale, i.e.\ the maximum wave-number, becomes:
\begin{equation}
\omega_\textrm{bn} = k_\textrm{max}\beta{c}.
\label{omega_bn_nonrel}  
\end{equation} 
Finally, a series of Lorentz transformations allow the generalization of these results to all velocities \citep{keenan15}. 

The introduction of a mean magnetic field will complicate this picture. The topic of radiation production by ultrarelativistic electrons in magnetized plasmas with small-scale magnetic fluctuations was originally considered in Ref. \citep{toptygin87}. In the case of strictly sub-Larmor-scale magnetic turbulence, with a mean field, the spectrum will simply be the sum of a synchrotron/cyclotron component (corresponding to the mean magnetic field) and the jitter contribution from the small-scale fluctuations, i.e.\ 
\begin{equation}
P(\omega) = P_\textrm{jitter}(\omega) + P_\textrm{synch}(\omega).
\label{mean_spec}  
\end{equation} 
Since a plasma is a dielectric medium, dispersion may affect the form of the radiation spectrum. The effect is mostly negligible in the ultrarelativistic limit, but dispersion may be required for a complete description of the mildly relativistic and non-relativistic regimes -- in real plasmas. Nonetheless, the dispersion-corrected spectrum has already been considered for small-angle jitter radiation, Ref. \citep{keenan15}, and synchrotron radiation \citep{zheleznyakov66}. For this reason, we will ignore plasma dispersion in this study.

When the electric field is stronger, or comparable, to the magnetic field, its contribution must be included. As shown by Ref.\ \citep{teraki14}, the radiation from ultrarelativistic particles in the ``small-scale'' regime resembles jitter radiation. At non-relativistic velocities, however, the deflection angle may be fairly large -- since the parallel acceleration on the electron cannot be neglected in this regime. Consequently, the radiation -- in the non-relativistic case -- may fall outside the small-angle jitter prescription.

Fortunately, since $\langle E^2 \rangle \ll \langle \delta{B}^2 \rangle$ for small-scale Whistler turbulence, we can completely ignore this electric contribution.

In Section \ref{s:results}, we will confirm Eq. (\ref{mean_spec}) (via ${\it ab\ initio}$ numerical simulation) in the case of small-scale (isotropic) Whistler turbulence. 

\section{Numerical model}
\label{s:model}

In Section \ref{s:analytic}, we made a number of theoretical predictions concerning the transport and radiation properties of magnetized plasmas with small-scale turbulent electromagnetic fluctuations. Additionally, we considered a concrete realization of this in the form of a cold, magnetized plasma embedded in sub-Larmor-scale Whistler turbulence. Here we describe the first-principle numerical simulations we employed to test our predictions.

As stated previously, we assumed the existence of a background of cold plasma which supports Whistler-mode turbulence. We then inject a smaller population of hot electrons (test particles) that are sub-Larmor-scale with respect to these preset Whistler magnetic fields. First, we consider the numerical generation of the Whistler magnetic and electric fields.

Our principal assumption, in generating electromagnetic turbulence, is that these stochastic electromagnetic fields are the linear superposition of a large number of wave-modes with randomized propagation direction and relative phase. Given this assumption, we can construct the turbulent fields directly from the plasma waves which are characteristic of the underlying instability. 

In general, the properties of these electromagnetic wave-modes, and their dispersion relation, are derived from the plasma dielectric tensor -- the determinant, of which, provides a system of characteristic equations. Given the``cold'' plasma approximation, these equations admit the dispersion relation specified by Eq. (\ref{whis_disp}) -- valid in the frequency range \citep{verkhoglyadova10}: 
\begin{equation}
\Omega_{ci} \ll \Omega_r \ll \Omega_{ce}
\label{freq_limits}
\end{equation}
where $\Omega_{ci} \equiv eB_0/m_ic$ is the ion cyclotron frequency and $m_i$ is an ion mass. The inequality is understood to hold for all ion species. The equations, additionally, specify the polarization of the wave-modes. Given obliquely (with respect to the ambient magnetic field) propagating whistler waves, the magnetic component will be right-circularly polarized with the following relations among its components \citep{sazhin93, verkhoglyadova10} 
\begin{equation}
\delta{B_x} = -\frac{1}{\tan(\theta_k)}\delta{B_z} = {\it i}\cos(\theta_k)\delta{B_y},
\label{b_polar_rel}
\end{equation}
where ${\bf B}_0$ is along the $z$-direction, and the wave-vector is in the $x$-$z$ plane. Because the magnetic field is divergenceless, ${\bf k} \perp \delta{\bf B}$. Given these conditions, the magnetic field will rotate about the direction of the wave-vector -- which, in the $\chi \ll 1$ regime, will have a period much greater than all other relevant time-scales. Since the phase is randomized for each wave-mode, this indicates that the magnetic field is approximately linearly polarized with a random polarization axis. 

Next, the electric field is (generally) elliptically polarized. It obeys the following relations \citep{sazhin93, verkhoglyadova10} :
\begin{equation}
\left\{\begin{array}{ll}
E_x/E_y = -{\it i}\Theta_1 \\
E_z/E_x = \Theta_2, \\
\end{array}\right.
\label{e_polar_rel}  
\end{equation} 
where 
\begin{equation}
\left\{\begin{array}{ll}
\Theta_1 \equiv\displaystyle{ \frac{k^2c^2\sin(\theta_k)\cos(\theta_k)}{\omega_{pe}^2 + k^2c^2\sin^2(\theta_k)}} \\[1.5em]
\Theta_2 \equiv \displaystyle{\frac{\Omega_r^2\omega_{pe}^2 + (\Omega_r^2-\Omega_{ce}^2)k^2c^2}{\Omega_r\omega_{pe}^2\Omega_{ce}}}. \\
\end{array}\right.
\label{Theta_defs}  
\end{equation} 
These equations suggest that the electric field parallel to the ambient magnetic field may be expressed in terms of the magnetic fluctuations via the relation \citep{sazhin93}:
\begin{equation}
|{\bf E}^{z}_{\bf k}| = \frac{\Omega_r^2}{\Omega_{ce}kc}|{\bf B}_{\bf k}|\tan(\theta_k).
\label{e_para_exact}  
\end{equation} 
Then, specifying a magnetic spectral distribution, e.g.\ Eq. (\ref{Bk}), allows the complete description of each wave-mode. We then add a large number of these waves (given random relative phases and ``k-vectors'') to simulate Whistler turbulence. 

Next, we describe the numerical solution of the equation of motion for our test electrons. Obviously, the test particles do not interact with each other, nor do they induce any fields. Additionally, any radiative energy losses are neglected. An individual electron's motion is, consequently, determined only by the Lorentz force equation given by:
\begin{equation}
\frac{d{\boldsymbol\beta}}{dt} = -\frac{1}{\gamma}\left[{\boldsymbol\Omega}_E +   {\boldsymbol\beta}\times{\boldsymbol\Omega}_B - {\boldsymbol\beta}\left({\boldsymbol\beta}\cdot{\boldsymbol\Omega}_E\right) \right],
\label{dvdt}
\end{equation}
where ${\boldsymbol\Omega}_E \equiv e{\bf E}/m_e{c}$ and $\boldsymbol\Omega_B \equiv \Omega_{ce}\hat{z} + \boldsymbol\Omega_{\delta{B}}$.

Eq. (\ref{dvdt}) was solved via a fixed step 4$^\text{th}$-order Runge-Kutta-Nystr\"om method, or a (symplectic) 2$^\text{nd}$-order Boris method. In our test runs, we found little variation between these two methods -- barring numerical instability due to using an insufficiently small step-size in time. This is likely because our simulation time is limited by actual computational time, and thus, we were unable to realize the slow accumulation of errors in the total energy characteristic of non-symplectic numerical integrators. 

With all the particle positions, velocities, and accelerations calculated, the numerical radiation spectrum was obtained directly from the Li\'{e}nard-Wiechert potentials. The radiation spectrum (which is the radiative spectral energy, $dW$ per unit frequency, $d\omega$, and per unit solid-angle, $d\eta$) seen by a distant observer is given by \citep{landau75, jackson99}:
\begin{equation}
\frac{d^2W}{d\omega\, d\eta} = 
\frac{e^2}{4\pi^2 c}  \left|\int_{-\infty}^\infty \! {\bf A}_{\boldsymbol\kappa}(t)e^{i\omega{t}}\, \mathrm{d} t
\right|^2,
\label{LW}  
\end{equation} 
where
\begin{equation}
{\bf A}_{\boldsymbol\kappa}(t) \equiv \frac{\hat{\bf n}\times[(\hat{\bf n} - {\boldsymbol\beta}) \times \dot{\boldsymbol\beta} ]}{(1 - \hat{\bf n}\cdot{\boldsymbol\beta})^2}e^{-i{\boldsymbol\kappa}\cdot {\bf r}(t)}.
\label{A_k}  
\end{equation} 
In this equation, ${\bf r}(t)$ is the particle's position at the retarded time $t$, $\boldsymbol\kappa \equiv \hat{\bf n}\omega/c$ is the wave vector which points along $\hat{\bf n}$ from ${\bf r}(t)$ to the observer and $\dot{\boldsymbol\beta} \equiv \text{d}{\boldsymbol\beta}/\text{d}t$. Since the observer is assumed to be distant, $\hat{\bf n}$ is approximated as fixed in time to the origin of the coordinate system. 

Next, the total radiation spectrum is obtained by ``summing'' over the spectra of the individual particles. For the moment, we will only consider mean-free, small-scale magnetic turbulence in the following discussion. 

Given an isotropically distributed (in velocity-space) ensemble of electrons, the ``summed'' spectrum will be equivalent to the angle-averaged, i.e.\ $dW/d\omega$, spectrum for a single electron. There are two, usually equivalent, methods for doing this ``summation''. First, one may add the spectra coherently by summing over each particle's ${\bf A}_{\boldsymbol\kappa}$, and then performing a single integration via Eq. (\ref{LW}). This method is more physical. Alternatively, we may add the spectra incoherently (i.e., by integrating each particle's ${\bf A}_{\boldsymbol\kappa}$ separately, and then summing the results of each integration). As discussed in Ref. \citep{hededal05}, both methods will result in the same spectra, since the wave phases are uncorrelated. However, an incoherent sum will produce a spectrum that is less noisy (for a given number of simulation particles) than the coherently summed spectrum. For this reason, we employ the incoherent approach in this study -- as we have done previously. 

In contrast to our previous studies, Refs. \citep{keenan13, keenan15}, the non-vanishing mean magnetic field introduces a previously non-existent anisotropy; the ``summed'' spectrum will, as a result, depend upon the location of the observer. However, if the magnetic turbulence is statistically homogeneous/isotropic, then the synchrotron/cyclotron (mean field) component of the spectrum, alone, will possess this dependence. Since the angle-averaged synchrotron spectrum is a known function, we may simply add it to the jitter spectrum, obtained via the ``summation'' method above. Lastly, the contribution due to the electric field may be neglected, since $\langle E^2 \rangle \ll \langle \delta{B}^2 \rangle$.

Finally, the electron pitch-angle (with respect to the $z$-axis) and kinetic energy, $W_e \equiv (\gamma - 1)m_ec^2$, were calculated at each time-step. Using the definitions, Eqs. (\ref{Daa_def}) and (\ref{Energy_diff_def}), we obtained the pitch-angle and energy diffusion coefficients directly from the simulation data.

\section{Numerical results}
\label{s:results}

In Section \ref{s:analytic}, we made a number of theoretical predictions concerning the transport and radiation properties of magnetized plasmas with small-scale turbulent electromagnetic fluctuations -- in particular, sub-Larmor-scale Whistler-modes. Additionally, we anticipated that an inter-connection between the transport and radiative properties of electrons moving through small-scale Whistler turbulence exists, as it does for strictly, ``Weibel-like'', mean-field-free turbulence \citep{keenan13, keenan15}.

\subsection{Whistler Turbulence}
\label{s:whist_results}

First of all, we explore the particle transport by testing our predictions concerning the energy and pitch-angle diffusion coefficients in small-scale Whistler turbulence. The diffusion coefficients depend on various parameters, cf. Eqs. (\ref{Daa_specific}) and (\ref{Dww_rel_daa}), namely the particle's velocity, $\beta$, the magnetic fluctuation field strength, $\langle \Omega_{\delta{B}}^2 \rangle$, and the field correlation scale, $\lambda_B$.

To start, we must confirm the fundamental assumption of diffusion. As stated previously, a diffusive process requires that both $\langle \Delta{W}_e^2 \rangle$ and $\langle \alpha^2 \rangle$ increase linearly in time -- at least, on some characteristic time-scale of the system. With $\delta{B}/B_0 \ll 1$, the ``gyro-period''
\begin{equation}
T_g \equiv \frac{2\pi\gamma}{\Omega_{ce}} = 2\pi\frac{\gamma{m_e}c}{eB_0},
\label{gyro_def}  
\end{equation} 
is such a characteristic, ``macro'' time-scale. On a multiple gyro-period time-scale, the electron velocities will change very slightly. Consequently, we may treat the magnitude of the electron velocity as approximately constant. 

\begin{figure}
\includegraphics[angle = 0, width = 1\columnwidth]{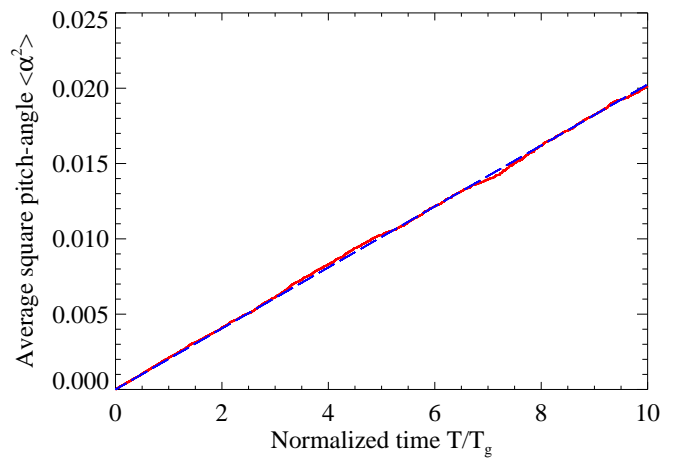}
\caption{(Color online) Average square pitch-angle vs. normalized time. Relevant parameters are $\beta = 0.25$, (number of simulation particles) $N_p = 5000$, $k_\text{min} = 32\pi$, $k_\text{max} = 10k_\text{kmin}$, $\langle \delta{B}^2 \rangle^{1/2}/B_0 = 0.1$, $\Omega_{ce} = 1$, $\rho \approx 400$, $\chi \approx 0.04$, and $\mu = 4$. The linear nature of the curve (solid, ``red'') confirms the diffusive nature of the pitch-angle transport. Here, the dashed (``blue'') line indicates a line of best fit (simple linear regression) with Pearson correlation coefficient: $0.9998$.}
\label{alph_diff}
\end{figure}

\begin{figure}
\includegraphics[angle = 0, width = 1\columnwidth]{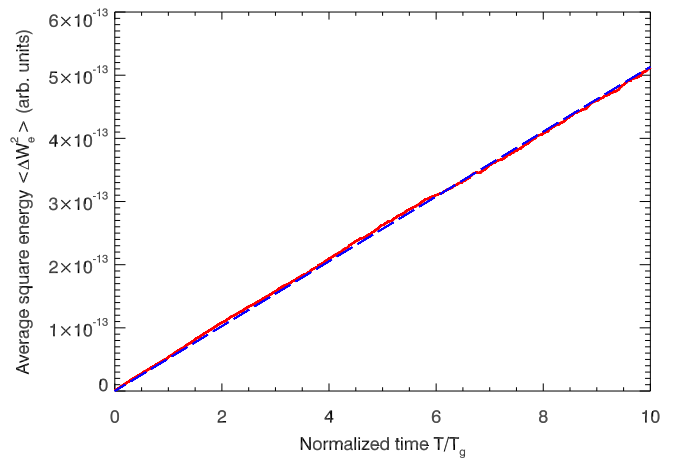}
\caption{(Color online) Average square change in electron energy (in simulation units) vs. normalized time. Relevant parameters are $\beta = 0.25$, (number of simulation particles) $N_p = 5000$, $k_\text{min} = 64\pi$, $k_\text{max} = 10k_\text{kmin}$, $\langle \delta{B}^2 \rangle^{1/2}/B_0 = 0.1$, $\Omega_{ce} = 1$, $\rho \approx 400$, $\chi \approx 0.04$, and $\mu = 4$. The linear nature of the curve (solid, ``red'') confirms the diffusive nature of the energy transport. Here, the dashed (``blue'') line indicates a line of best fit (simple linear regression) with Pearson correlation coefficient: $0.9999$.}
\label{erg_diff}
\end{figure}

\begin{figure}
\includegraphics[angle = 0, width = 1\columnwidth]{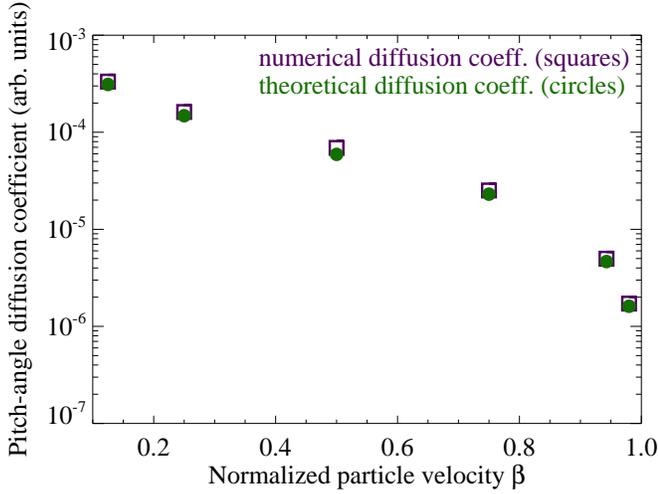}
\caption{(Color online)  Pitch-angle diffusion coefficient, $D_{\alpha\alpha}$ vs the normalized electron velocity, $\beta$. Relevant simulation parameters include: $N_p = 5000$, $k_\text{min} = 32\pi$, $k_\text{max} = 10k_\text{kmin}$, $\langle \delta{B}^2 \rangle^{1/2}/B_0 = 0.1$, $\Omega_{ce} = 1$, $\chi \approx 0.02$, and $\mu = 4$. The (purple) empty ``squares'' indicate the $D_{\alpha\alpha}$'s obtained directly from simulation data (as the slope of $\langle\alpha^2\rangle$ vs. time), while the (green) filled ``circles" are the analytical pitch-angle diffusion coefficients, given by Eq. (\ref{Daa_specific}). }
\label{v_alpha}
\end{figure}

\begin{figure}
\includegraphics[angle = 0, width = 1\columnwidth]{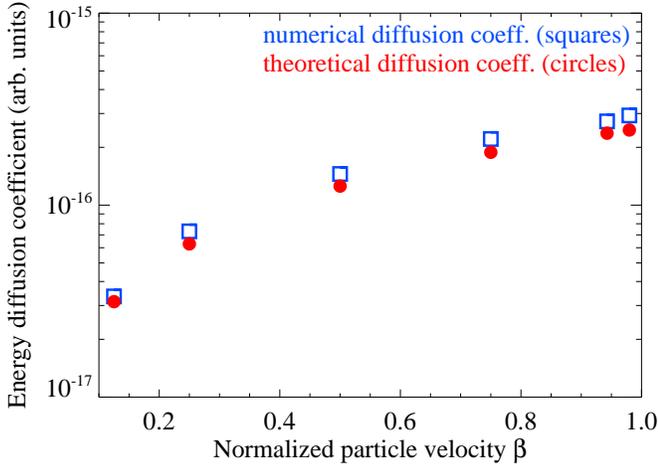}
\caption{(Color online)  Energy diffusion coefficient, $D_{WW}$ vs the normalized electron velocity, $\beta$. Relevant simulation parameters include: $N_p = 5000$, $k_\text{min} = 32\pi$, $k_\text{max} = 10k_\text{kmin}$, $\langle \delta{B}^2 \rangle^{1/2}/B_0 = 0.1$, $\Omega_{ce} = 1$, $\chi \approx 0.02$, and $\mu = 4$. The (blue) empty ``squares'' indicate the $D_{WW}$'s obtained directly from simulation (as the slope of $\langle\ \Delta{W}_e^2 \rangle$ vs. time), while the (red) filled ``circles" are the analytical energy diffusion coefficients, given by Eq. (\ref{Dww_rel_daa}). }
\label{v_erg}
\end{figure}

To establish diffusion, $5000$ mono-energetic electrons ($\beta = 0.25$) were injected into Whistler turbulence with $k_\text{min} = 32\pi$ (arbitrary simulation units), $k_\text{max} = 10k_\text{min}$, $\langle \delta{B}^2 \rangle^{1/2}/B_0 = 0.1$, $\Omega_{ce} = 1$, $\rho \approx 400$, $\chi \approx 0.04$, and $\mu = 4$. The simulation time included several gyroperiods; $T = 10T_g$. Additional simulation parameters include: the time-step $\Delta{t} = 0.00125$ (arbitrary units), and the number of Whistler wave-modes $N_m = 10000$. In Figure \ref{alph_diff}, we see that the average square pitch-angle (as measured with respect to the $z$-axis, i.e.\ the mean field direction) does, indeed, grow linearly with time. 
Figure \ref{erg_diff} confirms that the electron energy undergoes a classical diffusive process as well.
With the existence of pitch-angle and energy diffusion established, we then proceeded to compare the slope of $\langle \alpha^2 \rangle$ and $\langle \Delta{W}_e^2 \rangle$ vs time (the numerical pitch-angle and energy diffusion coefficients) to Eqs. (\ref{Daa_specific}) and (\ref{Dww_rel_daa}). In Figure \ref{v_alpha}, the numerically obtained pitch-angle diffusion coefficients are compared to Eq. (\ref{Daa_specific}) for a range of possible electron velocities. In each, the theoretical and numerical results differ only by a small factor of ${\cal O}(1)$.
Next, in Figure \ref{v_erg}, we see decent agreement with Eq. (\ref{Dww_rel_daa}) and the numerical energy diffusion coefficients. 
Figures \ref{v_alpha} and \ref{v_erg}, furthermore, confirm that our theoretical diffusion coefficients are valid for all electron velocities -- including relativistic speeds.  

\begin{figure}
\includegraphics[angle = 0, width = 1\columnwidth]{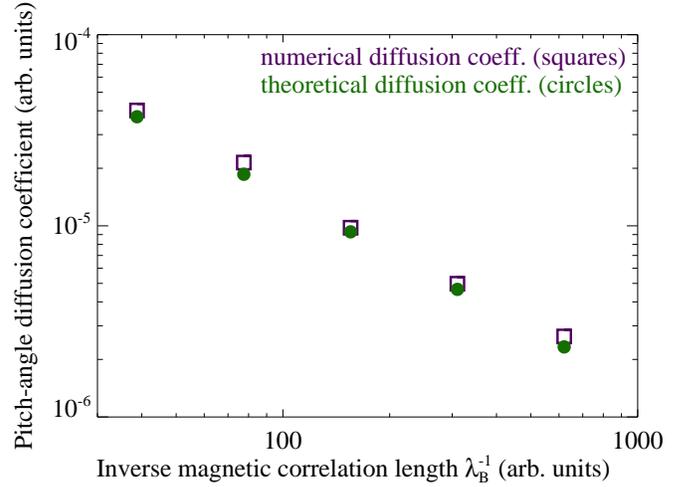}
\caption{(Color online)  Pitch-angle diffusion coefficient, $D_{\alpha\alpha}$ vs the inverse of magnetic field correlation scale, $\lambda_B^{-1}$. Relevant simulation parameters include: $\gamma = 3$, $N_p = 1000$, $k_\text{min} = 8\pi$, $16\pi$, $32\pi$, $64\pi$, and $128\pi$, $k_\text{max} = 10k_\text{kmin}$ (for each $ k_\text{kmin}$), $\langle \delta{B}^2 \rangle^{1/2}/B_0 = 0.1$, $\Omega_{ce} = 1$, $\chi \approx 0.02$, and $\mu = 4$. For each data point, the theoretical and numerical results differ only by a small factor of ${\cal O}(1)$. }
\label{alpha_corl}
\end{figure}

\begin{figure}
\includegraphics[angle = 0, width = 1\columnwidth]{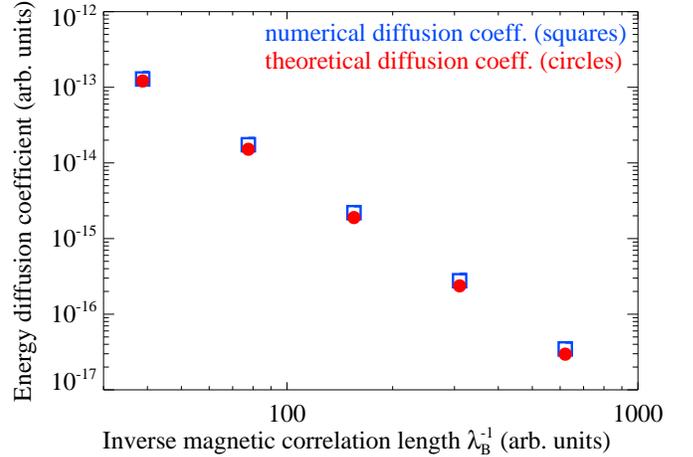}
\caption{(Color online)  Energy diffusion coefficient, $D_{WW}$ vs the inverse of magnetic field correlation scale, $\lambda_B^{-1}$. Relevant simulation parameters include: $\gamma = 3$, $N_p = 1000$, $k_\text{min} = 8\pi$, $16\pi$, $32\pi$, $64\pi$, and $128\pi$, $k_\text{max} = 10k_\text{kmin}$ (for each $ k_\text{kmin}$), $\langle \delta{B}^2 \rangle^{1/2}/B_0 = 0.1$, $\Omega_{ce} = 1$, $\chi \approx 0.02$, and $\mu = 4$. The theoretical and numerical results differ only by a small factor of ${\cal O}(1)$. }
\label{e_corl}
\end{figure}

Another important parameter which strongly influences the diffusive transport is the magnetic field correlation length. In Figure \ref{alpha_corl}, the correlation length was varied by changing $k_\text{min}$, while keeping all other parameters fixed.
Once more, we see close agreement with Eq. (\ref{Daa_specific}). Similarly, the numerical and theoretical energy diffusion coefficients continue to show decent agreement -- see Figure \ref{e_corl}. 

\begin{figure}
\includegraphics[angle = 0, width = 1\columnwidth]{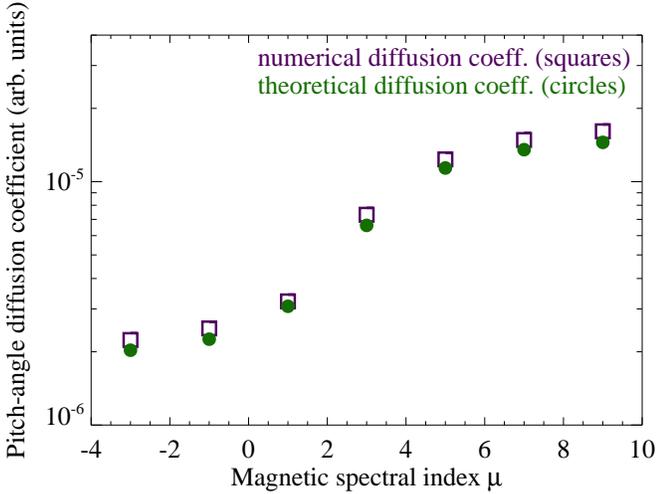}
\caption{(Color online) Pitch-angle diffusion coefficient, $D_{\alpha\alpha}$ vs the magnetic spectral index, $\mu$. Relevant parameters are $N_p = 2000$, $k_\text{min} = 32\pi$, $k_\text{max} = 10k_\text{max}$, $\langle \delta{B}^2 \rangle^{1/2}/B_0 = 0.1$, $\Omega_{ce} = 1$, and $\chi \approx 0.05$. Notice that the numerical results have nearly the same functional dependence on $\mu$ as the analytical squares, as given by Eq. \ref{Daa_specific}. }
\label{diff_mu_alpha}
\end{figure}

\begin{figure}
\includegraphics[angle = 0, width = 1\columnwidth]{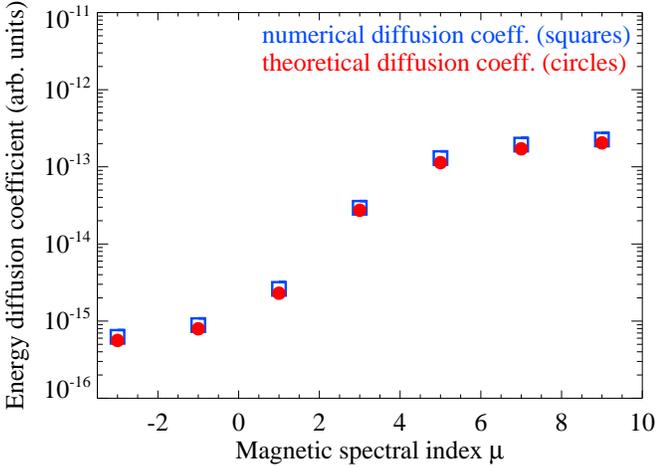}
\caption{(Color online) Energy diffusion coefficient, $D_{WW}$ vs the magnetic spectral index, $\mu$. Relevant parameters are $N_p = 2000$, $k_\text{min} = 32\pi$, $k_\text{max} = 10k_\text{max}$, $\langle \delta{B}^2 \rangle^{1/2}/B_0 = 0.1$, $\Omega_{ce} = 1$, and $\chi \approx 0.05$.}
\label{e_mu}
\end{figure}

Lastly, we consider the magnetic spectral index, $\mu$ -- i.e.\ the power-law exponent in Eq. (\ref{Bk}). With $k_\text{min} = 32\pi$ and $k_\text{max} = 10k_\text{min}$, we varied the magnetic spectral index, $\mu$ from $-3$ to $9$. In Figure \ref{diff_mu_alpha}, we see that the numerical pitch-angle diffusion coefficient closely matches the analytical result.
Similarly close agreement was, once again, realized between the energy diffusion coefficients; as may be seen in Figure \ref{e_mu}.

Finally, we consider the radiation spectra. As discussed in Section \ref{s:model}, the radiation spectra are expected to be the summation of synchrotron (cyclotron) and jitter (psuedo-cyclotron) components. For an ultrarelativistic electron, the angle-averaged synchrotron radiation spectrum is the known function \citep{landau75, jackson99}:
\begin{equation}
\frac{dW}{d\omega} = \sqrt{3}\frac{e^2}{c}\gamma\frac{\omega}{\omega_c}\int_{\omega/\omega_c}^\infty \!  K_{5/3}(x) \, \mathrm{d} x,
\label{synch_def}  
\end{equation} 
where $K_j(x)$ is a modified Bessel function of the second-kind, and $\omega_c = 3/2\gamma^2\Omega_{ce}$ is the critical synchrotron frequency. This result applies for an electron moving in the plane transverse to the ambient magnetic field, i.e.\ when $\alpha = \pi/2$. Nonetheless, we find the expression fits the synchotron components fairly well; especially when $\gamma$ is decently large.

\begin{figure}
\includegraphics[angle = 0, width = 1\columnwidth]{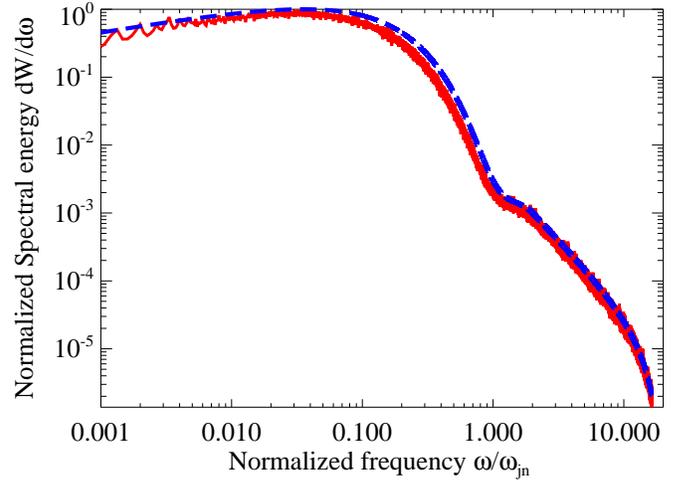}
\caption{(Color online) Radiation spectrum for a monoenergetic, isotropic distribution of $\gamma = 5$ ($\chi \sim 1$; $\rho \approx 928$; $\langle \delta{B}^2 \rangle^{1/2}/B_0 = 0.1$) electrons moving through small-scale Whistler turbulence. The frequency is normalized by $\omega_{jn} = \gamma^2k_\text{min}\beta{c}$ -- the relativistic jitter frequency. The solid (``red'') curve is from simulation data, whereas the dashed (``blue'') curve is the analytic estimate. Clearly, the spectrum is well represented by a superposition of synchrotron$+$jitter components. Note the lower-frequency synchrotron component and a higher-frequency power-law component corresponding to the small-angle jitter radiation.}
\label{spec_gam5}
\end{figure}

We illustrate two numerical spectra here, along with their corresponding analytical estimates -- for details concerning the jitter component, see Ref. \citep{keenan15}. First, we considered a $\gamma = 5$ electron population for Figure \ref{spec_gam5}. In this plot, the relevant parameters are: $N_p = 1000$, $\Delta{t} = 0.00125$, $k_\text{min} = 2\pi$, $k_\text{max} = 20\pi$, $\langle \delta{B}^2 \rangle^{1/2}/B_0 = 0.1$, $\Omega_{ce} = 0.512$, $\mu = 4$, $\rho \approx 928$, $\chi \sim 1$, and the total simulation time was $T = 5T_g$. We see that the synchrotron$+$jitter fit closely resembles the numerical spectrum.

\begin{figure}
\includegraphics[angle = 0, width = 1\columnwidth]{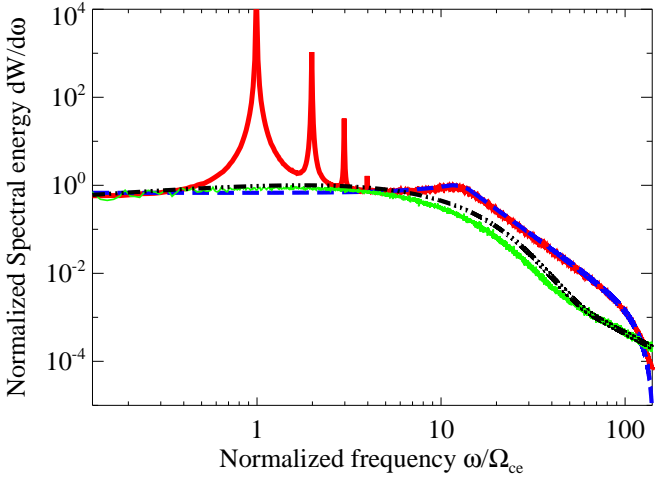}
\caption{(Color online) Radiation spectrum for a monoenergetic, isotropic distribution of $\beta = 0.125$ electrons ($\chi \sim 0.04$; $\rho \approx 160$; $\langle \delta{B}^2 \rangle^{1/2}/B_0 = 0.2$; $\Omega_{ce} = 2$; $k_\text{min} = 64\pi$; $k_\text{max} = 10k_\text{min}$; $\mu = 5$; $T = 50T_g$); superimposed with a spectrum given a population of $\gamma = 4$ electrons ($\chi \sim 1$; $\rho \approx 367$; $\langle \delta{B}^2 \rangle^{1/2}/B_0 = 0.1$; $\Omega_{ce} = 0.512$; $k_\text{min} = \pi$; $k_\text{max} = 10\pi$; $\mu = 4$; $T= 5T_g$). The normalization on the $y$-axis is arbitrary, whereas the $x$-axis is normalized to the $\beta = 0.125$ population's cyclotron frequency, i.e.\ $\Omega_{ce} = 2$. The ``thick'' solid (``red'') curve is from simulation data for the $\beta = 0.125$ population, the dashed (``blue'') curve is the corresponding analytic estimate for ``pure'' psuedo-cyclotron radiation, the ``thin'' solid line is the simulation data for the $\gamma = 4$ population, and the ``dot-dashed'' (	``black'') line is the $\gamma = 4$ analytic estimate. Notice, for the $\beta = 0.125$ spectrum, that the spectrum peaks near the cyclotron frequency, $\Omega_{ce}$ -- hence we see the signature of cyclotron radiation. The additional harmonics, which are purely a relativistic effect, are the signature of emerging synchrotron radiation.}
\label{cyc_syn_comp}
\end{figure}

Next, we explored the non-relativistic regime. In Figure \ref{cyc_syn_comp}, we assumed a population of sub-Larmor-scale $\beta = 0.125$ electrons. As expected, a peak in the spectrum may be observed near the cyclotron frequency $\Omega_{ce}$ -- confirming that the total spectrum is the hybrid of psuedo-cyclotron$+$cyclotron radiation. Additionally, to provide a point of comparision, we have superimposed a simulation result for $\gamma = 4$ electrons. 

\subsection{Langmuir Turbulence}
\label{s:lang_results}

In Section \ref{s:elec_pitch}, we predicted the pitch-angle diffusion coefficient for ultrarelativistic electrons moving in small-scale electric turbulence. Here, we will numerically confirm Eq.\ (\ref{daa_elec_def}) -- via our first-principle simulations. We will treat the electric field as quasi-static, i.e.\ ${\bf k}\times{\bf E}_{\bf k} \approx {\bf 0}$. To this end, we employ a model identical to that used by Ref. \citep{teraki14} for the numerical generation of the electrostatic (Langmuir) turbulence. Essentially, a background of ``cold'' langmuir wave-modes are assumed to be present, with $\Omega_r \sim \omega_\text{pe}$. 

It was assumed that the Langmuir oscillations are ``cold'', i.e.\ possessing real frequency, $\Omega_r({\bf k}) \approx \omega_\text{pe}$ (where $\omega_\text{pe}$ is the electron plasma frequency). In this case, the parameters which characterize the radiation/transport regime are the jitter parameter \citep{teraki14}:
\begin{equation}
\delta_j \equiv \frac{\delta\alpha_E}{\Delta\theta} \sim \frac{eE_{\perp}\lambda_E}{m_e c^2}
\label{jitter_para_def}
\end{equation}
and the ``skin-number'':
\begin{equation}
\chi \equiv \frac{d_e}{\lambda_E^t} = \frac{c}{\omega_\text{pe}\lambda_E^t}.
\label{skin_num_def}
\end{equation}
Additionally, we considered an electric field with a spectral distribution identical to Eq.\ (\ref{Bk}) -- with $\delta{\bf B}_{\bf k} \rightarrow {\bf E}_{\bf k}$. Furthermore, the simulation procedure was identical -- with the exception that ${\bf E} \parallel {\bf k}$, rather than peripendicular to the wave-vector.

This form of turbulence may be realized in a number of ways. ``Cold'' Whistler turbulence with $\chi \gg 1$ -- i.e.\ the opposite regime to that considered in the previous sections -- has an the electric field which is approximately electrostatic; i.e.\ resembling an anisotropic realization of Langmuir turbulence (ignoring the magnetic field), with $\Omega_r({\bf k}) \approx \Omega_\text{ce}cos(\theta_k)$. For strictly sub-Larmor-scale magnetic fields, the correlation length transit time is much shorter than the average gyro-period -- hence the electric field is effectively time-independent. Conceptually, the electric field may be comparable in strength to the magnetic field in this regime. Consequently, it may be necessary to include its contribution.

\begin{figure}
\includegraphics[angle = 0, width = 1\columnwidth]{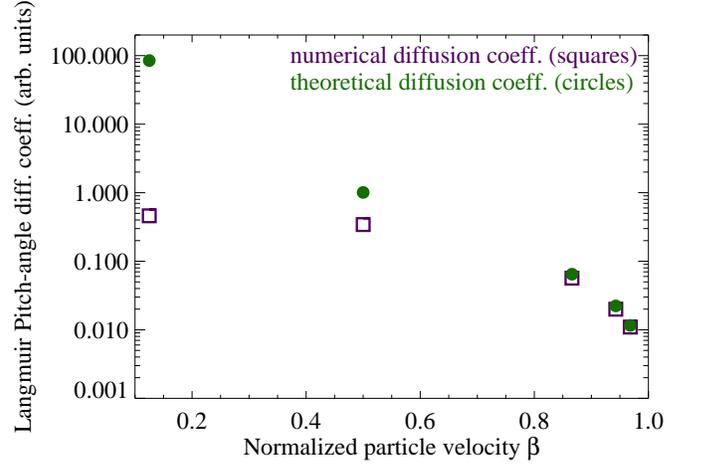}
\caption{(Color online) Electric pitch-angle diffusion coefficient, $D^\text{elec.}_{\alpha\alpha}$ vs the normalized electron velocity, $\beta$ for small-scale Langmuir turbulence. Relevant simulation parameters include: $k_\text{min} = 8\pi$, $k_\text{max} = 10k_\text{kmin}$, $\langle \Omega_E^2 \rangle = 4.0$, $\chi \approx 666.67$, $\delta_j \approx 0.08$, and $\mu = 5$. The (purple) empty ``squares'' indicate the $D^\text{lang.}_{\alpha\alpha}$'s obtained directly from simulation data, while the (green) filled ``circles" are the analytical pitch-angle diffusion coefficients, given by Eq. (\ref{daa_elec_def}). Notice that the small deflection approximation, which is the foundational assumption behind  Eq. (\ref{daa_elec_def}), holds well for velocities that are mildly relativistic ($\gamma \sim 2$). }
\label{diff_coeff}
\end{figure}

\begin{figure}
\includegraphics[angle = 0, width = 1\columnwidth]{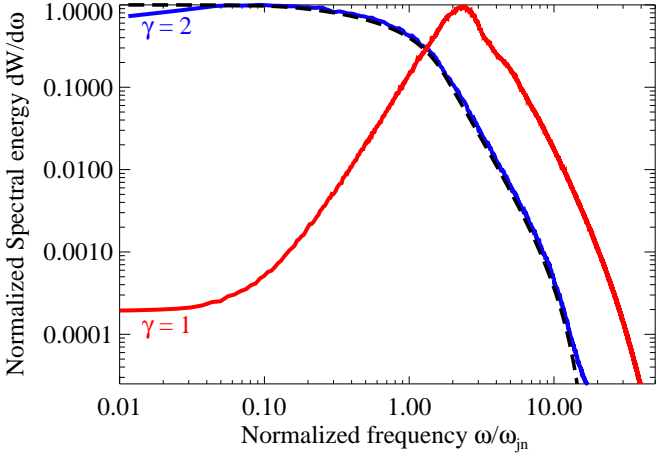}
\caption{(Color online) Langmuir Radiation spectra for the $\gamma = 2$ and $v = 0.125c$ electrons (see Figure \ref{diff_coeff} for details on the simulation parameters). The frequency is normalized by the characteristic jitter frequency, i.e.\ $\omega_{jn} \equiv \gamma^2k_\text{min}\beta{c}$. The lower (``red'') curve is from simulation data, and it corresponds to the $v = 0.125c$ electron. The upper (``blue'') curve is the simulation result for the $\gamma = 2$ electron, and the dashed curve is the analytic estimate. Clearly, the mildly relativistic spectrum is well represented by the (Langmuir) jitter result.}
\label{spect}
\end{figure}

Figure \ref{diff_coeff} shows the electric pitch-angle diffusion coefficient as a function of particle velocity. In each scenario, $10000$ monoenergetic electrons were injected into Langmuir turbulence with $\delta_j \approx 0.08$, $\chi \approx 666.67$, $k_\text{min} = 8\pi$, $k_\text{max} = 10k_\text{min}$, and $\langle \Omega_E^2 \rangle = 4.0$ (all simulation units are arbitrary). The electron velocities vary for each run. We see that the numerical pitch-angle diffusion coefficient approaches the ultrarelativistic result as $v \rightarrow c$. Furthermore, we see fairly close agreement, even in the mildly relativistic ($\gamma \sim 2$) regime. The large discrepancy seen from the most leftward data points may be attributed to the breakdown of the small deflection angle approximation, which accompanies the existence of a comparable longitudinal acceleration.
In Figure \ref{spect}, we have plotted the corresponding numerical radiation spectra (the spectral energy per unit frequency) for electrons with $v = 0.125c$ and $\gamma = 2$. Details on the numerical implementation may be found in Refs. \citep{keenan13, keenan15}. We present the analytical solution for the $\gamma = 2$ electron via the perturbation theory approach detailed in Ref. \citep{keenan15}. The resulting radiation spectrum is analogous to the (mildly) relativistic small-angle jitter spectrum of an electron moving through sub-Larmor-scale magnetic turbulence, but it is morphologically distinct. This is because the electrostatic field, owing to its curl-free presentation, has a different correlation tensor, $\Phi_{ij}({\bf k})$, than the (divergenceless) magnetic equivalent. Thus, we require the substitution:
\begin{equation}
\Phi_{ij}({\bf k}) \propto \left(\delta_{ij} - \hat{k}_i\hat{k}_j\right) \rightarrow \hat{k}_i\hat{k}_j.
\label{ten_sub}
\end{equation} 
The analytical solution, strictly, holds for the ultrarelativistic limit. Nonetheless, as can be seen in Figure \ref{spect}, the numerical solution closely matches the analytic result for mildly relativistic electrons with $\gamma = 2$. This is consistent with the result seen in Figure \ref{diff_coeff}, which suggests the presence of the small deflection angle regime. 

In contrast, the third spectrum in Figure \ref{spect} differs markedly from the analytic (jitter) prediction. This is the spectrum resolved for a $v = 0.125c$, i.e.\ $\gamma \approx 1$, electron. In accord with Figure \ref{diff_coeff}, the deflection angle is large, thus the spectrum is outside the small-angle jitter regime.

It is noteworthy that the $\chi \gg 1$ condition in Langmuir-like turbulence may not be physically realizable, since Landau damping would likely eliminate wave-modes at sub-skin-depth spatial scales too quickly \citep{teraki14}. With the field variability time-scales of comparable order to the electric correlation length transit time, it may be necessary to consider the rms electric field as a function of time. Thus, a more realistic model may require a time-dependent pitch-angle diffusion coefficient. 

\section{The Jitter/Synchrotron Spectrum of a Thermal Distribution of Particles}
\label{s:thermal}

In most cases, our sub-Larmor-scale electron distribution will not be composed of mono-energetic electrons. Here, we consider the radiation spectrum one might expect from a Maxwell-Boltzmann (thermal) distribution of electrons in sub-Larmor-scale magnetic fields.

To obtain the jitter component of the spectrum, we must average the single electron spectrum over an appropriate relativistic Maxwell-Boltzmann distribution. We define the jitter {\it emission coefficient}, which is the total radiated power per frequency per volume, as thusly:
\begin{equation}
\left(\frac{dP}{d\nu{dV}}\right)^\text{jitt.} = n_e\frac{\int \!  \ P_j(\nu, p)e^{\gamma/\Theta} \, \mathrm{d}^3p}{\int \!  \ e^{\gamma/\Theta} \, \mathrm{d}^3p},
\label{jitter_emission}
\end{equation}
where $\Theta \equiv k_BT_e/m_ec^2$, $k_B$ is the Boltzmann constant, $T_e$ is the electron temperature, $n_e$ is the electron number density, $\nu = \omega/2\pi$, and
\begin{equation}
P_j(\nu, p) \equiv \frac{2\pi}{T}\frac{dW}{d\omega}(p),
\label{P_j}
\end{equation}
is the single electron (power) spectrum with kinetic momentum, $p = \gamma m_ev$, and at the observation time, $T$. 

Next, we require an expression for the angle-averaged thermal synchrotron emission coefficient. To that end, we employ:
\begin{equation}
\left(\frac{dP}{d\omega{dV}}\right)^\text{syn.} = \frac{2^{1/6}\pi^{3/2}e^2n_e\nu}{3^{5/6}cK_2(1/\Theta)\xi^{1/6}}\text{exp}\left[-\left(\frac{9\xi}{2}\right)^{1/3}\right],
\label{synch_emission}
\end{equation}
where $\xi \equiv \omega/\Omega_\text{ce}\Theta^2$. This expression produces the correct total power, up to a factor of $1.05$, when $\Theta = 0.6$ \citep{wardzi00}. With $\Theta = 0.6$, the thermal Lorentz factor, $\gamma_{T_e} = \Theta + 1 = 1.6$. Thus, this corresponds to the trans-relativistic regime. 

When the temperature approaches the ultrarelativistic limit, i.e., $\Theta \gg 1$, Eq.\ (\ref{synch_emission}) gives a fairly accurate result, with a correction factor of order unity ($\approx 0.744$ -- see Ref.\ \citep{wardzi00}, for details).
\begin{figure}
\includegraphics[angle = 0, width = 1\columnwidth]{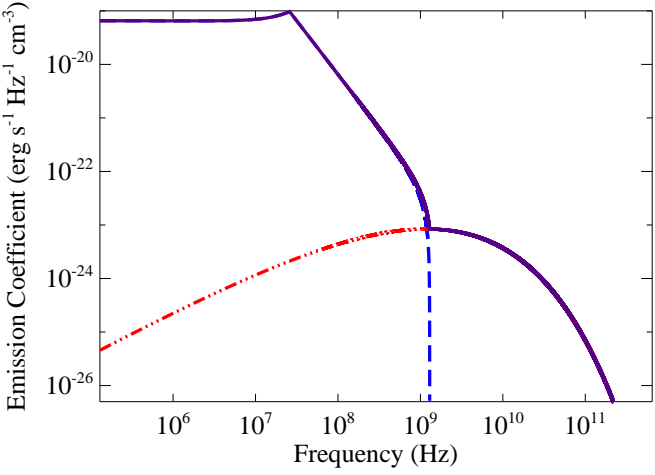}
\caption{(Color online) Emission coefficient vs. frequency for a thermal distribution of electrons moving through ``magnetized'', sub-Larmor-scale magnetic turbulence. Relevant parameters: $n_e = 1 \ cm^{-3}$, $\delta{B} = B_0 = 1 \ G$, $\gamma_{T_e} = 12$, $k_\text{max} = 50k_\text{min}$, and $k_\text{min}^{-1} = d_e^\text{rel.}$ -- where $d_e^\text{rel.} \equiv c\sqrt{\gamma_{T_e}}/\omega_\text{pe}$ is the relativistic skin-depth. The jitter component --  dashed (``blue'') line -- overpowers the synchrotron portion -- three-dot-dashed (``red'') line -- at frequencies below $\omega_\text{bn} \sim \gamma_{T_e}^2k_\text{max}v_{T_e}$. This produces a distinctly non-synchrotron feature, at low frequencies, in the total (summed) spectrum, solid (``purple'') line. }
\label{gam12}
\end{figure}
\newline
\indent
In figure \ref{gam12}, we have plotted the combined emission coefficient for a scenario in which sub-Larmor-scale magnetic turbulence, with a spectrum identical to Eq.\ (\ref{Bk}), is embedded in an ambient magnetic field, ${\bf B}_0$. We suppose the following parameters: $n_e = 1 \ cm^{-3}$, $\delta{B} = B_0 = 1 \ G$, $\gamma_{T_e} = 12$, $k_\text{max} = 50k_\text{min}$, and $k_\text{min}^{-1} = d_e^\text{rel.}$ -- where $d_e^\text{rel.} \equiv c\sqrt{\gamma_{T_e}}/\omega_\text{pe}$ is the relativistic skin-depth. These parameters, other than $\Theta$, are not important for determining the overall shape of the spectra; thus, our selection is made only for instructional purposes.
\begin{figure}
\includegraphics[angle = 0, width = 1\columnwidth]{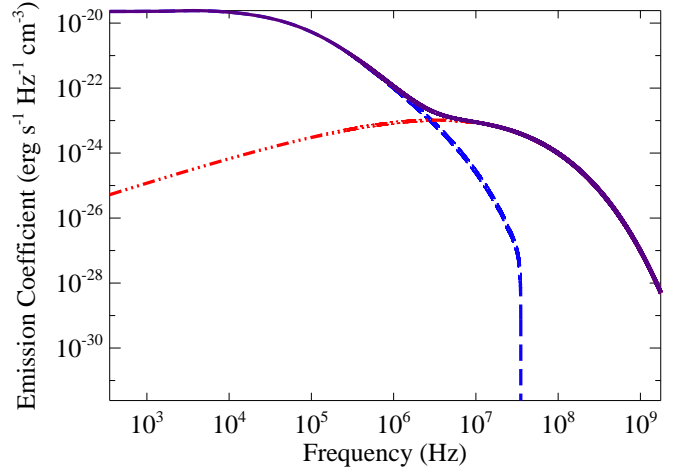}
\caption{(Color online). Emission coefficient vs. frequency for a thermal distribution of electrons moving through ``magnetized'', sub-Larmor-scale magnetic turbulence. Relevant parameters: $n_e = 1 \ cm^{-3}$, $\delta{B} = B_0 = 1 \ G$, $\gamma_{T_e} = 1.6$, $k_\text{max} = 50k_\text{min}$, and $k_\text{min}^{-1} = d_e^\text{rel.}$. Despite the presence of noticable thermal spread, the jitter component --  dashed (``blue'') line -- still overpowers the synchrotron portion -- three-dot-dashed (``red'') line -- at frequencies below $\omega_\text{bn} \sim \gamma_{T_e}^2k_\text{max}v_{T_e}$. The summed spectrum, solid (``purple'') line, remains distinctly non-synchrotron-like at low frequencies.}
\label{gam06}
\end{figure}
\newline
\indent
As may be readily seen in Figure \ref{gam12}, the jitter emission spectrum -- dashed (``blue'') line -- dominates over the synchrotron component -- three-dot-dashed (``red'') line -- at low frequencies. This contrasts with the scenario depicted in Figure \ref{spec_gam5}, where the jitter portion dominates at high frequencies. Essentially, the ratio: $\omega_\text{jn}/\omega_\text{c}$, determines where the jitter component makes an appearance.

Furthermore, the depicted jitter and synchrotron spectra are nearly identical to the mono-energetic equivalents. This is because with $\gamma_{T_e}  = 12$ -- or, equivalently, $\Theta = 11$ -- the vast majority of particles are moving ultrarelativistically. Hence, the thermal spread is very small.

In contrast, with $\gamma_{T_e} = 1.6$, a considerable degree of thermal spread will be noticable on inspection. However, as we see in Figure \ref{gam06} -- where we consider an identical scenario, with $\Theta = 0.6$ -- this spread does not obscure the trans-relativistic jitter (pseudo-cyclotron) feature; the jitter portion is still very clearly distinct from the thermal synchotron component.

To summarize, the signature of jitter radiation --- both in the relativistic and trans-relativistic regimes -- remains clearly evident, even given a thermal distribution of electrons.

\FloatBarrier

\section{Conclusions}
\label{s:concl}

In this paper, we explored test particle transport (diffusion) and radiation production in magnetized plasmas with small-scale electromagnetic turbulence. In our previous works \citep{keenan13, keenan15}, we showed that the pitch-angle diffusion coefficient and the simultaneously produced radiation spectrum are wholly determined by the particle velocity and the statistical/spectral properties of small-scale (mean-free) magnetic turbulence. Here, we have generalized these results to the case when the magnetic field has a mean value.

In fact, the pitch-angle diffusion coefficient, Eq. (\ref{Daa_specific}), remains unchanged by the addition of a mean field -- so long as the pitch-angle, $\alpha$ assumes its conventional meaning, i.e.\ as the angle between the electron velocity vector and the ambient (mean) magnetic field. Since magnetized plasmas characterized by instability often include random electric fields, as is the case for the Whistler turbulence considered here, we additionally considered test particle energy diffusion. We showed that the energy diffusion coefficient in small-scale Whistler turbulence is directly proportional to the (magnetic) pitch-angle diffusion coefficient -- see Eq. (\ref{Dww_rel_daa}). Thus, it is also intimately related to the field's statistical properties. Consequently, transport via energy diffusion may provide, yet another, powerful diagnostic tool for the investigation of small-scale electromagnetic fluctuations in magnetized plasmas.

Whistler turbulence, as conceived here, is dominated by the magnetic field. In constrast, the purely electrostatic Langmuir turbulence is characterized by random electric fields. We showed that a generalization of the magnetic pitch-angle diffusion coefficient exists for the case of relativistic electrons moving through small-scale electric turbulence. We, further, confirmed our analytic result via first-principle numerical simulations of Langmuir turbulence.

Next, we showed that the test particle radiation spectrum (which is predominately determined by the magnetic field in Whistler turbulence) is simply the summation of a small-scale, jitter/pseudo-cyclotron component and a regular, synchrotron/cyclotron component -- see Eq. (\ref{mean_spec}). We have, further, confirmed these theoretical results via first-principle numerical simulations. 

Additionally, we confirmed the result first shown in Ref. \citep{teraki14} that the spectrum of relativistic electrons in small-scale Langmuir turbulence is a form of jitter radiation. We, further, expanded upon this result by resolving the spectrum for trans-relativistic velocities -- showing that the jitter prescription holds well even down to $\gamma \sim 2$.

Finally, we considered the radiation produced by a Maxwell-Boltzmann (thermal) distribution of electrons in a magnetized plasma with sub-Larmor-scale magnetic fluctuations. We demonstrated that the signature of the jitter component clearly remains when the fluctuation field is comparable to the ambient magnetic field -- just as it did for the mono-energetic case considered previously.

Our model implicitly considered a scenario whereby a turbulent magnetic field was generated in a cold, magnetized, background plasma. We then imagined the existence of a ``hot'' population of sub-Larmor-scale electrons that served as our test particles. We suggested that the motion of high-energy, supra-thermal, ``super-Halo'' electrons through the magnetized solar wind is a promising candidate for the physical realization of our model. Indeed, despite the fact that this population only accounts for a small fraction of a percent of the solar wind, its high energy ($2-20 \ keV$) makes it very significant \citep{wang15, yoon15}.

Additionally, the super-Halo population is largely insenstive to solar activity, and it is likely constantly present in the interplanetary plasma \citep{wang15} -- thus, it is a relatively fixed source of high-energy particles. In fact, recent work has suggested that the super-Halo electrons may mediate Weibel-like instabilities in the solar wind plasma -- facilitating the development of  Kinetic-Alfv{\' e}n wave (KAW) and/or Whistler-mode turbulence at sub-electron spatial scales \citep{che14}.

The nature of this wave turbulence, in the solar wind plasma, is a matter of contention. Conficting accounts implicate either KAW or Whistler-modes (or both) \citep{mithaiwala12, salem12}. A number of reasons for this ambiguity have been given. For example, in situ measurements of these waves must be done in the spacecraft frame -- which is usually moving at super-Alfv{\`e}nic speeds with respect to the plasma \citep{salem12}. Furthermore, the solar wind hosts a permanent source of turbulence; hence, many results implicating Whistler waves -- via, for example, the observed power spectrum -- may be the erroneous signature of the, ever present, background turbulence \citep{lacombe14}.

However, a more detailed analysis of the turbulent spectrum may provide a means by which Whistlers and KAW may be distinguished. In fact, the degree of anisotropy has been found to significantly differ between the two types of wave turbulence \citep{salem12}. With regard to our model, the presence of anisotropy will result in diffusion coefficients that differ perpendicular and parallel to the anisotropy axis (which is typically the direction of the ambient magnetic field), since the correlation lengths will depend upon the structure of the correlation tensor. 

Hence, we may imagine that the transport properties of ``hot'' electrons (e.g. sub-Larmor-scale, super-Halo electrons) may be different for Whistler-mode and KAW turbulence. The radiation spectrum would, additionally, distinguish these forms of turbulence -- as the anisotropy, which features into the field correlation tensor, would alter the shape of the radiation spectrum in a characteristic way. 

Other cases where this work is of great interest include the upstream of collisionless shocks in astrophysical and interplanetary systems. The ``hot'' population, in this case, would be Cosmic Rays (CRs) -- which are both non-relativistic and relativistic in astrophysical systems. Relativistic CRs are radiatively efficient and radiation from them is observed in supernova remnant shocks (Tycho, Chandra, 1003, etc) pulsar wind nebulae, termination shocks, GRBs (internal and reverse shocks, if the ejecta is magnetized) and GRB remnants. In the latter case, the external shock may become weak and non-relativistic. Consequently, the ambient interstellar field may become significant, and Whistler-like instabilities may develop from an initial Weibel ``seed''.

Concerning Whistler turbulence and our energy diffusion coefficient, our model's principal limitation is the essential assumption of the ``cold'' plasma approximation. In many cases, thermal effects must be accounted for; i.e.\ the plasma ``beta'' is non-negligible. Nonetheless, under certain conditions, the underlying plasma may be considered ``cold''. As an example, the plasma outflow in ultrarelativistic ``collisionless'' shocks is beam-like, with very little dispersion; this permits a cold plasma treatment \citep{lemoine10}. Therefore, since these shocks may be mediated in part by small-scale Whistler-modes, our rough estimates concerning the diffusive transport of electrons may provide some insight into the process of shock acceleration. 

To conclude, the obtained results reveal strong inter-relation of transport and radiative properties of plasmas turbulent at sub-Larmor scales -- ``magnetized'', i.e.\ possessing a mean magnetic field, or otherwise.

\begin{acknowledgments}

This work was partially supported by the DOE grant DE-FG02-07ER54940 and the NSF grant AST-1209665.

\end{acknowledgments}

\end{document}